# Constructive Axiomatics in Spacetime Physics Part III: A Constructive Axiomatic Approach to Quantum Spacetime

Emily Adlam,* Niels Linnemann† and James Read‡

**Abstract**

The Ehlers-Pirani-Schild (EPS) constructive axiomatisation of general relativity, published in 1972, purports to build up the kinematical structure of that theory from only axioms which have indubitable empirical content. It is, therefore, of profound significance both to the epistemology and to the metaphysics of spacetime theories. In this article, we consider extensions of the EPS axiomatisation towards quantum general relativity based upon quantum mechanical inputs (Part III). There are two companion papers, in which we provide a pedagogical walkthrough to the EPS axiomatisation (Part I), and discuss the significance of constructive approaches to spacetime structure more generally (Part II).

## Contents

*eadlam2@uwo.ca
†niels.linnemann@uni-bremen.de
‡james.read@philosophy.ox.ac.uk

# 1 Introduction

In 1972, Ehlers, Pirani, and Schild (henceforth EPS) presented an axiomatisation of (the kinematics of) the theory of general relativity (GR) which purported to build up the spacetime structure of that theory from only (supposedly) indubitable empirical posits regarding light rays and particle trajectories (Ehlers et al., 2012). Leaving aside the extent to which the scheme strictly fulfils its constructivist ambitions of non-circularly constructing spacetime structure from basic observational statements—and leaving out in particular the extent to which EPS managed to provide a scheme for interpreting GR without posits of external clocks (a stated intention of their article, which was after all published as part of a *Festschrift* for the chronometer Synge)—the EPS scheme and its subsequent amendments constitute invaluable tools for assessing classical theory space, including the necessity and/or sufficiency of GR to account for a certain body of empirical data.

In this paper, we aim to understand the extent to which something resembling the EPS approach to constructing spacetime can be applied when the inputs are quantum mechanical rather than classical (as in the original EPS construction). That is, we consider the EPS axiomatic approach to GR's spacetime structure with all classical light ray signals replaced by quantum light signals, and all particle signals replaced by quantum particles; this can be done, or so we will argue, in a natural fashion (indeed, a number of different natural fashions).[1] In making these substitutions and applying the EPS approach, one ultimately derives a superposition of metric structures as the relevant kinematical structure for quantum spacetime; moreover, as we will see, there is a way of interpreting these outputs in terms of branching spacetime structures.

---

[1] Notably, Audretsch and Lämmerzahl (1991) have extended the EPS scheme by considering matter that is explicitly modelled as the classical limit of quantum matter (postulate 1 and 2), and even as rays superposed to wave packages (postulate 2′). However, the possibility for the geometric background to become superposed in virtue of such superposition is not considerd.

In more detail, what will the resulting kinematics look like? First off, it seems likely that we will not thereby arrive at a single Hausdorff manifold. This follows directly from a standard result of differential geometry: for a Hausdorff manifold $M$, a geodesic beginning at a point $p$ in $M$ with an initial tangent vector $x$ must be unique for a non-zero length of time $t$. This fails to be true in our case, because the expectation is that light rays will participate in spatial superpositions—for example, in a beam-splitter a light ray may be placed in a superposition of two different paths. Assuming that no spacetime point exists in more than one branch (we will return to this point), the two geodesics must lie along different points, so the geodesic from the divergence point fails to be unique.[2]

Secondly, in the quantum case, it is clear that axioms can't be *wholly* based upon empirical observations, as is purportedly the case in the original EPS construction. This is because in the case where light enters a superposition, we can't directly observe the light going down both paths (either because doing so would (effectively) collapse the wavefunction, or because in an Everettian picture observers are confined to individual branches of the wavefunction), so the existence of the superpositions is an inference that we make to explain the observed interference effects, rather than a direct observation.[3] Moreover, we also expect that massive particles will participate in superpositions and as a result we will get superpositions of different spacetime structures (whether at the level of the metric or of the manifold—we will return to this). Once again, these superpositions will not be observable directly, and in fact at present we do not even have indirect evidence for them, since no experimental demonstration of the existence of spacetime superpositions has yet been obtained—the existence of spacetime superpositions is largely a theoretical conjecture at this stage.[4] So the axioms that we employ in our quantum EPS construction will be somewhat conjectural, meaning that we have a certain amount of freedom about how to select them. The methodology we employ here will be based on keeping the axioms as close as possible to the original EPS axioms while reflecting standard principles concerning the behaviour of light and particles in a quantum world.

Now, one might ask why it is useful to carry out an EPS construction when the original operational motivations no longer apply. However, there is more than one way to motivate the EPS construction. One could see it—arguably as EPS themselves would have done—as a minimal empirically-driven construction demonstrating that the kinematics of GR are the correct ones based on the existing evidence. Alternatively, one could take an essentially relationalist position which insists that the structure of spacetime is literally built up out of the behaviour of light rays and particles, and therefore if light rays and particles necessarily behave as specified in the EPS axioms, then it follows that spacetime necessarily has the structure that EPS derive. The EPS construction is then to be understood not merely as a *epistemic justification* for the relativistic kinematics, but at the same time seen as a metaphysical *explanation*, i.e., grounding for it.[5] But once we are aiming for explanation *rather* than epistemic justification in the quantum case, we are perfectly entitled to even use axioms describing behaviour that can't be directly observed—obviously the plausibility of the resulting explanation will depend on the *a priori* plausibility of the axioms, but then that is a common feature of scientific explanations. Indeed from this point of view it's quite remarkable that the original EPS construction apparently succeeds in explaining the nature of the relativistic kinematics based solely an axioms which can be verified by direct empirical observations (though it should be noted that there are disagreements about the extent to which it does indeed succeed—see Linnemann and Read (2021a)). Moreover, in addition to the idea that one can use EPS to realise a relationalist vision with respect to the *metrical* structure of spacetime by grounding this in the existence and behaviour of light rays and particles, we note that if the radar coordinate construction is taken seriously and is understood to pertain to the behaviour of actual (and not merely hypothetical) light rays and particles, then it can be regarded as an implementation of a certain kind of structuralism (cf. Esfeld and Lam (2008)) whereby spacetime points have their identities only in virtue of the behaviour of matter. The EPS construction can then be regarded as evidence that, at least at the level of local kinematics, GR is compatible with such versions of relationalism/structuralism, provided that we supply ourselves with a sufficient number of light rays and particles. Attempting a similar construction in the quantum case thus offers interesting insight into how the relationalist/structuralist might fare in a quantum gravitational context and whether there are any additional challenges over and above those faced in classical GR.

---

[2]Later, on this issue, following e.g. Luc and Placek (2020), we'll distinguish between 'local' branching (in which curves 'split'), and 'global' branching (in which curves do not 'split', but the spacetime in question has 'trousers').

[3]Admittedly, already in classical EPS one explicitly renders light and particle detection as indirect in the sense that one only measures the effect of light and particles rays on probing light signals in terms of the light signals' reception time. Quantum EPS' operational procedure of comparing light ray input to (interference) output is thus at least in this way continuous to the operational protocol in classical EPS.

[4]That said, recently-proposed experiments in table top quantum gravity in reach of actual implementation are seen by many to have the potential to change this. See Huggett et al. (2022) for a critical review.

In addition, quantum EPS has significance beyond metaphysics: attempting to perform a quantum EPS construction is a highly educational exercise that confronts one with a number of questions regarding the nature and consequences of superpositions of spacetimes, with important lessons for quantum gravity research, in particular the right quantisation of GR. Thus this construction offers a different perspective on some of the conceptual challenges associated with quantum gravity, where we arrive at these problems from a top-down operational perspective rather than a bottom-up perspective based on some specific theoretical framework. Any bottom-up approach to quantum gravity must ultimately take a stance on the questions we will address, either by building a methodology based on certain answers to these questions or by deriving answers to these questions from its basic methodology. For example, the first question we have to answer in a quantum EPS construction is 'Can there exist superpositions of different spacetime structures?' and of course all mainstream approaches to quantum gravity (excluding semiclassical gravity) answer this question in the affirmative. The EPS approach offers us the opportunity to parse questions of this kind and examine the consequences of various possible answers to them in a way that is independent of any particular theoretical approach to quantum gravity.

---

[5] For a similar case where the epistemic basis also serves as the metaphysical (or modal) basis, see Adlam (2022). Forms of Humeanism with an 'epistemic basis' arguably also are examples of this kind.

We proceed as follows. §2 rehearses the core features of classical EPS (for a more detailed introduction, see Linnemann and Read (2021a)). In §3, we consider the **core obstacle** to quantum EPS: how to relate spacetime structure across superpositions? The discussion leads to various **variants** on how to undertake a quantum EPS project, which will be presented in more detail in §4. In §5, we select two variants and exhibit some possible axioms for a quantum EPS scheme. Finally, in §6, we draw the following major **lessons** from this work:

- *Physical lesson:* From a strict relationalist point of view, spacetime superpositions may not be so conceptually distinct from spatial superpositions.

- *Theory construction lesson:* There is an extreme ambiguity as to which structure is subject to quantisation upon quantising a theory like GR.

- *Metaphysical lesson on relationism:* Quantum EPS can be understood as a relationist/structuralist project for quantum spacetime. (Related to this, we present in this article some reflections on how our work bears upon the hole argument of GR.)

## 2 Core features of Classical EPS

Classical EPS, as presented in Ehlers et al. (2012), is characterised by three core attributes:

- *Constructivist* in the sense of (Carnap, 1967, §2) (note that Carnap uses the word 'constructional' where we use 'constructivist'):

  > ... to construct *a* out of *b, c* means to produce a general rule that indicates for each individual case how a statement about *a* must be transformed in order to yield a statement about *b, c*. This rule of translation we call a construction rule or constructional definition (it has the form of a definition; cf. §38). By a constructional system we mean a step-by-step ordering of objects in such a way that the objects of each level are constructed from those of the lower levels. Because of the transitivity of reducibility, all objects of the constructional system are thus indirectly constructed from objects of the first level. These *basic objects* form the *basis* of the system.

- *Constructive* in the sense of Reichenbach (1969): The basis consists of immediately empirically accessible objects or quantities. Arguably, this also implies that statements are of local nature (locality).

- *Kinematical*: The (generalised) scheme is concerned only with setting up a kinematical space for GR (spacetime theories more generally), not its dynamics.[6]

We will see that it may not be possible for a quantum version of EPS to maintain all of these principles; in particular, it may be necessary to replace constructive elements by structures less amenable to direct empirical access. However, we reiterate that the classical EPS construction also diverges from these principles at times—in particular in setting up the basic manifold structure but also at various other steps when tacit assumptions are smuggled in (see Linnemann and Read (2021a), in particular §8.1).

## 3 Obstacles to Quantum EPS

The central challenge for a quantum version of EPS has, we contend, to do with the notion of spacetime point identity. In this section, we present and discuss the issues involved here, as well as review a range of manners in which such issues might be tackled.

### 3.1 Point identities in classical manifolds

The starting point of the EPS construction is a set of spacetime points which are initially postulated as distinct individuals despite the fact that at this stage there is no physical structure to distinguish them.[7] If these mathematical spacetime points are regarded as representing real physical spacetime points endowed with primitive identities, then it appears to deliver us a kinematics which would potentially be vulnerable to the hole argument (although of course in order to actually make the hole argument we also have to add dynamics to the theory)—we will return to this issue later.

However, in fact this starting assumption is arguably harmless, because once the manifold has been constructed by appeal to the behaviour of particles and light rays, we may lift the assumption of primitive identities by supposing that points are identified only by the particles and light rays that pass through them. We may then think of the actual set of events as referring to the domain of *physical* radar coordinates—and physical radar coordinates alone: each point of a physical radar coordinate neighbourhood is a (possible) crossing of light rays. In particular, a change between such physical radar coordinates can be understood as inducing only a trivial spacetime diffeomorphism, i.e. a diffeomorphism that does nothing (rather than just leaving the equations invariant).[8]

---

[6]See (Curiel, 2016) for an apt account of the kinematical/dynamical distinction, and Linnemann and Read (2021b) for some further discussion.

[7]A constructivist program *à la* EPS is restrictionist in nature: notions like 'event' are initially defined sparsely (and thus permissively), and only then the notion gets more and more restricted through the introduction of empirically-motivated structure. In this restrictionist sense, the EPS programme is in the spirit of Klein's Erlangen approach to geometry—for a modern discussion, see Wallace (2019).

In this manner, it is reasonable to regard EPS' use of spacetime points as a convenient mathematical stepping stone. In fact, the EPS construction can therefore be regarded as an example of Cao's notion of 'self-consistent bootstrapping' (Cao, 2001), which he advocates as a solution to the problem of how we can understand the gravitational field as ontologically prior to the manifold 'spacetime', despite the fact that the latter is usually required to define local degrees of freedom to begin with (i.e., the gravitational field is formulated as a field on the manifold).[9] The idea is that we start with "a non-physical bare manifold, on which parameter localisation can be tentatively defined," then build up our theory on that manifold, and finally argue that "[i]f the final results in a diffeomorphism covariant theory are independent of any specificity of parameter localisation, except for some most general features of the gravitational field ... then the whole procedure is a justifiable way of investigating the physical spacetime and its ontological underpinnings." EPS can be read as following a similar approach, except that instead of starting from a manifold they begin with a set of bare points and derive the manifold structure therefrom: since the final construction does not require one to identify spacetime points across different possible worlds, there is no longer any need for these points to have primitive identities, so the bootstrapping has succeeded.

## 3.2 Point identities in manifold superpositions

In constructing the kinematics for a theory of quantum gravity, the primitive identity question has additional complexity due to the possibility that we will get superpositions of spacetime structures, which raises the question of whether we can identify points across branches. A naïve substantivalist about spacetime would perhaps want to insist that points have primitive identities which allow for a unique identification of points across branches—even though the actual identification still has to be chosen and arguably well-justified. In contrast, her structuralist counterparts who take it that spacetime points have their identities only in virtue of the behaviour of light and matter in their vicinity will presumably deny the possibility of such a straightforward identification map. Now, those falling into the latter of these camps may still attempt to employ the bootstrapping method advocated by Cao—assuming primitive identities for points at first and subsequently lifting that assumption—but it's less clear that this will succeed in the quantum context. The reason we were always free to make such a move in the classical case was that the kinematics produced by the EPS construction involves no internal modal notions, i.e. within a representational context each model represents a single possible world and its interpretation does not require reference to any other possible world, and therefore the construction will clearly never require any claims about identities of spacetime points across different possible worlds represented by one kinematical model; on the other hand, the kinematics produced by the quantum EPS construction will presumably involve distinct branches of the wavefunction each containing some spacetime points, and therefore it's possible that the construction will involve claims about identities of spacetime points across different possible branches and in this sense worlds represented by one kinematical model. So if we are not willing to accept primitive identities of spacetime points, we will have to decide how to deal with this.

---

[8] This is not EPS' own reading of the radar coordinates; EPS themselves employ radar coordinate charts as part of the usual manifold atlas. This means that they will be linked to non-trivial diffeomorphism and thus, provided the dynamics is diffeomorphism-invariant, will be subject to hole argument-type objections as well.

[9] A similar issue has been discussed in the context of the dynamical approach to spacetime theories promulgated by Brown and Pooley (see Brown (2005); Brown and Pooley (2001, 2006))—see Norton (2008); Menon (2019); Chen and Fritz (2021); Linnemann and Salimkhani (2021).

Can we simply refrain from postulating any identities between points in different branches? This is likely to lead to problems if we ultimately expect to end up with kinematics *directly* suitable for a theory like low energy perturbative quantum gravity.[10] In particular, we'd expect to be able to make sense of effects like the Bose-Marletto-Vedral (BMV) experiment (Bose et al. (2017); Marletto and Vedral (2017)), where we have two massive particles in spatial superpositions, giving rise to four branches of the wavefunction. Because the masses are in different positions in each branch, and the masses are sources of gravitational fields, it follows from general relativity's identification of the gravitational field with the structure of spacetime that we must have different spacetimes in every branch, but of course at the end the masses are all supposed to end up together in the same spacetime again, having picked up different phases due to time dilation. But if there is no natural map between the four distinct spacetimes, it's unclear how the particles in the separate branches could possibly be brought back together to recombine, or how we could identify the points in each branch where they are supposed to meet. It would seem that once particles or photons go into different spacetimes they should never be able to interact again—they now live on different manifolds so there's no sensible way to define something like a scalar product between the correponding wavefunctions of three-metrics embedded in different manifolds.

We also need to consider the possibility of what Anandan (1997) dubs 'quantum diffeomorphisms', which are individual diffeomorphisms applied separately within different branches of the wavefunction (and the associated manifolds)—so that one may have, for example, a quantum diffeomorphism which acts as the identity in one branch, but non-trivially in another (this will be relevant in our discussions of the hole argument below).[11] Just as we might demand that a classical spacetime theory be invariant under diffeomorphisms, so it would seem natural (by a generalisation of Einstein's principle of gen-eral covariance) to expect that quantum spacetime theories be invariant under quantum diffeomorphisms. But the possibility of quantum diffeomorphisms will cause difficulties if we wish to understand phenomena like recombination and interference as local effects: for an interaction between particles in different branches that is local in one choice of coordinate system will cease to be local under quantum diffeomorphisms. This suggests that if we don't have identities between points in different branches we can't appeal directly to locality to understand when and where phenomena like recombination happen, which makes it a little hard to see how such a thing could possibly happen at all.

---

[10]We discuss in the following sections the degree to which the physical setup of the dynamics actually has to be sensibly constrained by the kinematical setup.

[11]We discuss further quantum diffeomorphisms and their relation to general covariance in §A.

Indeed, even if the BMV prediction and other predictions of low-energy perturbative quantum gravity are wrong, there are still problems, because the issue goes beyond particular scenarios involving quantum gravity phenomenology: once we believe that some spacetime superpositions are possible, it follows that any time a massive object is in a spatial superposition, no matter how small the mass, the spacetime in which it lives must split into two separate spacetimes. Of course, for the small masses we deal with in current quantum experiments the difference between the spacetimes is experimentally insignificant and thus it is typically assumed that we can completely discount any effects of gravitational back-reaction. But if we accept that there are no identities between spacetime points in different branches of a spacetime superposition, it would seem to follow that no matter how small the difference is, once objects enter these different spacetimes they will never again be able to interact: once a massive particle enters a spatial superposition, it should never be possible to recombine the branches. But of course we frequently perform experiments where masses enter spatial superpositions and are then recombined—for example, this occurs every time we put a massive particle into an interferometer where it enters a superposition of two different paths, with subsequent interference between these branches demonstrating that recombination has indeed occurred. So even based on existing experimental evidence we know that if spacetime superpositions are possible it must be possible for their branches to recombine, at least in the case where the differences between the two spacetimes are small.

Moreover, if we have total freedom to perform arbitrary quantum diffeomorphisms, then there's also the worry that in many relevant cases we will usually be able to perform separate diffeomorphisms in each branch until the spacetime structure looks the same in all the branches. For example, Anandan (1997) considers a case in which we have superpositions of different spacetime structures around a 'cosmic string' which has a superposition of different angular momenta. For each of the branches of this superposition we can find a gauge in which the metric is flat in each simply connected region outside the string, and yet a neutron approaching the string will undergo intensity oscillations as a result of this superposition, whereas it would not oscillate if it were simply in a classical region with a flat metric. So the superposition must be taken seriously as an element of reality even though it can apparently be transformed away by a quantum diffeomorphisms. This demonstrates that if we want to set up a theory of quantum gravity in such a way that there is freedom to perform arbitrary quantum diffeomorphisms, these diffeomorphisms must be implemented very carefully in order to ensure that the physical predictions are preserved under diffeomorphisms even when the differences between spacetime structure can be transformed away. In Anandan's case, the solution is to pay attention to the way the Hamiltonian constraints transform under the diffeomorphism: the same commutation relations are obtained before and after the diffeomorphism and thus the physical predictions are the same. More generally, in principle we would expect that even if a superposition state of manifolds can be brought into a form such that the metric is alike on all of them, the initial difference in metric structure will have been transferred into the matter sector: there should be no way to make metric and matter sector look the same, if the initial differences in metric structure have a substantive physical effect.[12]

These considerations suggest strongly that the theory we're building towards will ultimately need a way to identify at least some points across different branches. Of course, we don't necessarily have to implement such a thing at the level of the kinematics: as already alluded to above, it could be introduced dynamically. That the space of kinematical possibilities vastly underdetermines the range of dynamical possibilities lies in its very nature—the kinematical space is after all the arena relative to which to define the dynamics. Moreover, in the formulation of a theory there is often some freedom to shift constraints between the dynamics and the kinematics. For example, if we want our theory to be invariant under the transformation which adds $2\pi$ to some quantity $\theta$, we can either set up our kinematics such that $\theta$ is defined on a circle and hence is periodic (kinematical symmetry), or we can set up a very general kinematics such that $\theta$ can take any value on the real line and then demand through the dynamics that all observables are periodic functions of $\theta$ (dynamical symmetry). Spekkens (2015) in particular has argued that as a consequence of this freedom of choice, the distinction between kinematics and dynamics is not empirically grounded and should give way to other paradigms; this, indeed, is also part of the moral of the 'dynamical approach' to spacetime theories (see Brown (2005)). On the other hand, this somewhat conventionalist approach to the kinematic/dynamic distinction may be opposed by the viewpoint of kinematical statements as constitutive (and thus, in a sense, non-empirical) statements only on top of which dynamical statements can be formulated to begin with (see Curiel (2016)). And, surely, even if there is some vagueness around the question of where to draw the line between the kinematical and the dynamical will be vague, there may still be value in using the distinction to characterise physical theorising.

For example, one possible approach to physical theorising involves creating an extremely general kinematics which allows for all sorts of unphysical constructions, and then imposing *any* expected symmetries and invariances at the level of the dynamics only. In the quantum gravity case, constructing the most general kinematics would likely lead us to something that looks like an approach spearheaded by Hardy (2019), i.e. a bundle of manifolds with maps be-tween them which ultimately are to be regarded as being devoid of physical significance. We would then impose invariance under quantum diffeomorphisms as a constraint on the dynamics, and also use the dynamics to model 'splitting' and 'recombining' by imposing laws with the consequence that a given pair of manifolds has identical matter content except inside some (not necessarily connected) region, so the manifolds can be regarded as one and the same except inside that region. So the kinematics would postulate many distinct manifolds; identity between (parts of) these manifolds would be fixed by the dynamics. In any case, a manifold-based approach has the advantage that the construction of the kinematics can be done entirely locally via an EPS-like construction, leaving non-local features of the theory like holonomies to be encoded in the dynamics.

---

[12]This observation reinforces that the issue of linking up 'branching spacetimes' is not just about relating metric structure but rather about relating events as encoded by all field contents together.

However, there may be reasons to prefer a different approach. For a start, we note that in classical theories the notion of 'local' is kinematical rather than dynamical—that is to say, although questions about whether the interactions allowed by the theory are local must await the specification of some dynamics, the question of what counts as 'local' in the first place is settled at the level of kinematics. So one might naturally think that in the quantum case too it should be necessary to determine what counts as 'local' for interactions between branches before defining any dynamics. Moreover, one might hope that coming up with a kinematics which more closely reflects the symmetries of the dynamics might ultimately make it easier to actually come up with the correct dynamics. For example, arguably the main point of difference between loop quantum gravity and quantum geometrodynamics/quantum general relativity is a different choice of classical kinematics—loops versus 3-metrics/4-metrics—and thus far the loop approach has concretely been more successful; so there is an argument to be made that paying attention to these sorts of issues during the construction of the kinematics may be preferable to deferring it all to the dynamics. A less general kinematics for the quantum gravity case might employ something like a non-Hausdorff (branching) manifold, thus building the possibility of splitting and recombination into the kinematics rather than the dynamics. However, as we would require some non-local structures to define the region of branching and recombining, it might not be possible to construct such a thing using the EPS methodology alone.

We also note that if one accepts that the local construction of EPS leads only to a kinematical space of superpositions of manifolds but does not itself relate points across manifold structures, then one has to show that there can be well-defined interactions for various kinds of maps between the various superposed manifold structures in some other sense. In fact, the kinematical space seems for instance still sufficient to write out regularities and find a best-system relative to them without a *presupposed* standard of identifying physical structures across the manifolds. (That one can indeed find a well-defined and satisfactory best system, however, needs to be demonstrated. It might be helpful to draw on an analogy to Huggett's regularity-relationalism (Huggett, 2006) and, more generally, super-Humeanism (Esfeld and Deckert, 2017).)

So, let us consider now in more specific detail the central options for dealing with the problem of identifying points across branches. In brief, we take the different options to consist in the following:

1. Prevent significant spacetime superpositions from forming at all.

2. Identify points across branches based on a criterion of similarity.

3. Identify points across branches by stipulation, potentially independently of physical goings-on.

Examples of each of these strategies have appeared in the literature, so we will now examine them in turn.

### 3.2.1 No large superpositions: Penrose's collapse approach

Penrose (1996) advocates option (1) above. He invokes the equivalence principle, which he interprets as telling us that "it is the notion of free fall which is locally defined," and thus he contends that, if we want to map one spacetime onto another, the natural way to do that is to insist that their geodesics map onto one another, at least locally. Penrose then observes that for general superpositions of spacetime structure it will not be possible to find such map, so the prospects for a natural standard of cross-branch identity don't seem promising. Moreover, Penrose considers the absence of such a map to be a serious problem, because it blocks the construction of a global time-translation operator, so we have no way of defining the states of well-defined energy (which are usually defined as the eignestates of the time-translation operator). Inspired by the standard theory of unstable particles, Penrose takes the absence of a well-defined energy to mean that the state is unstable, with a lifetime of $\hbar$ divided by an appropriate measure of the energy uncertainty.[13] He therefore argues that as soon as superpositions become large enough to produce interestingly different spacetime structure in each branch, they will become unstable and undergo a gravitationally-induced collapse. Thus, in Penrose's scheme, ultimately we will never have to deal with the question of how to map points across branches, because as soon as the superposition becomes large enough that there is no longer a natural way to do this, it will immediately collapse.

### 3.2.2 Dynamical identification: Barbour's best-matching approach

Barbour (2001), on the other hand, feels that Penrose "is trying to solve a problem that has already been solved." The solution, according to Barbour, is option (2) above: specifically, we can define cross-branch identities using the very same best-matching procedure that Barbour has already developed to explain how three-dimensional slices are put together into a four-dimensional object in his 'Machian relationalist' alternatives to Newtonian mechanics and general relativity (see Mercati (2018) for a recent book-length review). This is done by defining a measure of difference (e.g. the difference between the values of the 3-metric at matching points, integrated over all points) and seeking a set of identities between points such that the global measure of difference is extremal. The configuration which extremizes this measure of difference is the best match. Moreover, Barbour has argued that "the three momentum constraints really do express the guts of Einsteinian dynamics and show that it arises through the creation of a metric on superspace by best-matching comparison of slightly different 3-geometries." (p. 211) Given that GR already appears to make use of best matching in this sense, Barbour argues that it makes sense to use best matching in quantum gravity as well: that is, in Barbour's view the problem of finding identities between points across different branches of the wavefunction is exactly the same as the problem of tracking 'the same point' across time in general relativity.

---

[13]One might wonder about what extent a global time evolution is to be expected in a theory of quantum gravity. Arguably, Penrose's posit could, however, be read charitably as claiming that an approximate standard of global time evolution has to become available at some point towards lower energies (albeit not necessarily in the deep quantum gravity regime).

That said, there does seem to be a difference between these two applications of best matching. In the case of matching points across time, arguably nothing actually depends on the best matching identification of points, because all our current field theories obey a locality requirement which ensures that fields only interact when they are co-located, and hence there can be no interactions between fields at spacetime points in different time slices. Thus although the best matched construction makes everything look simpler, in principle we could use a more complicated construction based on some arbitrary choice for how to identify points across spacetimes which would predict the same empirical results. So the door is open to interpret best matching as a conventionalist fact about the way in which we experience and construct spacetime and formulate our laws, rather than a tool which is used by nature itself. On the other hand—and as we have already seen—in the branching case the identification of points across branches would seem to have real empirical consequences if we agree that objects in different branches are supposed to interact and the branches recombine, with the interactions and recombinations being local, i.e. taking place at 'the same points' in each branch. Assuming that we're not willing to let go of that locality assumption, it would seem that using different identities across branches would produce different predictions, since we would end up with different combinations of fields interacting. Consider for illustration the BMV experiment in which two masses in superposition states are taken to get entangled with each other through gravitational interaction (and gravitational interaction alone): the relevant literature tacitly assumes that the location of the masses are all relative to one joint lab frame—no matter whether the experiment is modelled through a Newtonian potential, or (low-energy) metric fields, as done by Christodoulou and Rovelli (2019). But if we perform a quantum diffeomorphism which shifts the point at which the recombination occurs within one of the branches, then the phase change will be different and the interference effects will change. So if we use best matching to determine the point of interaction across different branches we are no longer free to interpret it as a fact about the way in which we ourselves construct spacetime: nature itself must make use of best matching.

Moreover, Barbour's best matching approach looks difficult to implement within the EPS scheme for two reasons: (i) Barbour's best matching requires us to compare the whole of a three-geometry all at once, and therefore we don't have access to that kind of procedure if we are purely dealing with local kinematics. And (ii) best matching requires us to employ the actual metric, so it most likely has to be implemented at the level of the dynamics. So if we accept Barbour's approach to cross-branch identities, it would seem that we have to give up on the ideas that 'locality' can be defined at the level of kinematics as well as that it is a meaningful concept to the local observer herself!

A possible approach to option (2), i.e., dynamical identification, other than that of best matching would be to invoke an effective field theory description. Here, we split the metric into a constant background $\eta_{\mu\nu}$ plus some small fluctuations $h_{\mu\nu}$, where $\eta_{\mu\nu}$ is constant across all the branches so only the fluctuations $h_{\mu\nu}$ are quantized. We can then use $\eta_{\mu\nu}$ to define a 'background' which tells us where the particles recombine. Several options are then available. We could adopt a description similar to that suggested by Penrose, where spacetime superpositions can occur provided that the difference between the spacetimes is small enough that the EFT description is possible, but once the differences become too large for this description to work, the superposition must collapse. Alternatively, we could say that spacetime superpositions can only be recombined as long as the differences are small enough for the EFT description to be valid; once the branches become too different it's no longer possible for them to be recombined, so we get a decoherence-like effect and the branches become separate non-interacting Everettian worlds. However, both of these options seem a little troubling, because even when the difference between the spacetimes is small, the identification between the spacetimes given by $\eta_{uv}$ is only an approximate matter, and yet it would seem that the resulting identity map which determines where the recombination occurs can't be an approximate matter. Also, it seems odd that the possibility of recombination should depend upon how similar the branches are, because recombination itself doesn't seem to come in degrees: either the branches come back together into one spacetime or they split and subsequently never interact, so there isn't really any intermediate option. Thus the EFT route forces us to postulate a very sharp transition between 'sufficiently similar' and 'not sufficiently similar,' even though similarity between branches is something that presumably varies along some sort of continuous scale. Moreover, these options would seem to compel us to espouse either a collapse model or the Everett interpretation, so the EFT approach may not be appealing if we want to avoid collapses and we also want to avoid multiple worlds.

### 3.2.3 Kinematical indifference: Hardy's q-diffeomorphism approach

Hardy (2019) takes up option (3) in the context of his work on quantum coordinate systems: he advocates a 'gauge description', by which is meant that we arbitrarily choose some way of mapping points from one branch to another. This enables him to arrive at a scheme of 'quantum manifolds', in which each branch of the wavefunction corresponds to a distinct manifold. (Note that Hardy does not prove that the quantum manifolds are in fact manifolds in the usual technical sense—but potentially the EPS scheme could be used to do that.) The coordinate systems on these manifolds define a mapping between them but the mapping is purely conventional and nothing should ultimately depend on the choice of map. It follows that the final theory should indeed be invariant with respect to quantum diffeomorphisms, which (recall) are diffeomorphisms whose action can be different within each branch of the wavefunction.[14]

This means that if we do take up option (3), we will have to insist that all solutions be invariant under quantum diffeomorphisms (let's call these 'q-diffeomorphisms' in what follows). In particular, if identities between spacetime points in different branches are indeed to be regarded as unphysical—as a mere 'gauge choice'—one might wonder whether our standard techniques for dealing with systems exhibiting gauge freedom can be extended to this context. Most gauge theories of interest are associated with constrained Hamiltonian systems, where the presymplectic phase space $N$ of the physical theory arises as a regular submanifold of a symplectic geometry. In the case of canonical general relativity, $N$ is defined by a set of four constraint functions per space point; the three momentum or vector constraints, which enforce diffeomorphism invariance, and the Hamiltonian or scalar constraint, which enforces invariance with respect to global time translations. When we quantize the theory, in order to maintain gauge invariance, we turn each first-class classical constraint into a quantum operator $C$ and insist that the space of physical states must obey the constraint $C|\psi\rangle = 0$—i.e., the constraints still vanish on the physical Hilbert space. Let us now take as an example the geometrodynamical approach to quantum gravity which treats the 3-metric $h_{\mu\nu}(x)$ as a field defined on a manifold and then quantizes it to give $\hat{h}_{\mu\nu}(x)$. The field eigenstates $|e\rangle$ are the states $\hat{h}_{\mu\nu}(x)|e\rangle = h_{\mu\nu}(x)|e\rangle$, so each eigenstate describes a configuration across spacetime. We can define a wave functional $\Psi[e] = \langle e|\Psi\rangle$, i.e. the wave functional gives the amplitude for any field configuration, which is to say that in general it describes a superposition of field configurations. Due to the diffeomorphism constraint, $\Psi[e]$ is invariant under coordinate transformations in 3-space, i.e. the same amplitude is assigned to a configuration and to its shifted version. Thus, as long as different branches in the superposition of spacetime structure are completely independent and non-interacting, the wave functionals and the theory more broadly are invariant under the action of q-diffeomorphisms.

However, it's unclear that the branches are in fact non-interacting: as noted above, typically it is expected that we should be able to see interference effects between different branches of the wave functional. Because interference between distinct branches is not possible within the classical theory, the classical constraints aren't designed to enforce invariance of interactions between branches under q-diffeomorphisms, so it is not obviously the case that the quantised versions of these constraints will successfully enforce invariance with respect to q-diffeomorphisms. Indeed, it can be shown that Dirac quantisation, in which we first quantize the theory and then apply the constraints, is equivalent to reduced quantisation,[15] in which we first apply the constraints and then quantize the theory, at least in the one-loop approximation. Clearly if we use reduced quantisation the constraints cannot yield covariance with respect to quantum diffeomorphisms, because the structure that makes quantum diffeomorphisms possible was not available prior to quantisation. Thus it seems likely that Dirac quantisation also will not yield invariance with respect to quantum diffeomorphisms, so there remains a worry that the standard constraint quantisation used in the construction of quantum theories of gravity will not in fact always remove all the gauge degrees of freedom.

---

[14]Anandan (1997) makes roughly the same point, suggesting that invariance under quantum diffeomorphisms can be regarded as a quantum principle of general covariance. Giacomini et al. (2019) discuss a similar issue within the 'quantum reference frame' research programme, although their quantum reference frames are defined at a given time, whereas Hardy works with coordinate systems for entire spacetimes.

Further difficulties arise as soon as we include interactions with other fields. All the important field theories of modern physics are local, i.e. they postulate only interactions of the form $\phi(x)\chi(x)$ defined between two fields at the same point. So in the quantum field theory we'd expect to see coupling terms of the form $h_{\mu\nu}(x)\phi(x)$ where $\phi(x)$ is some matter field coupled to the metric. Again, this will work so long as we're happy for $\phi(x)$ to split up into distinct branches with each branch coupled only to the metric inside its individual branch and no interactions between different branches, but in reality we expect the matter field $\phi(x)$ to exhibit interference between its branches, and since interference is a local phenomenon it would seem that performing diffeomorphisms inside one branch and not another might change the resulting interference of the matter field $\phi(x)$. In order to avoid this we would presumably need to require that the q-diffeomorphisms go smoothly to zero at all points where matter fields exhibit self-interactions, and yet we can't enforce that requirement without first having a standard of cross-branch identity of points, since otherwise we will have no idea which fields $\phi(x), \phi(x')$ we are supposed to match up.

One possible way to write down a kinematics for quantum gravity such that it will indeed have its simplest expression in a form which is covariant with respect to q-diffeomorphisms may be to write it in terms of closed loops, or holonomies, since closed loops are preserved by diffeomorphisms.[16] This of course is exactly what quantum loop gravity does. Because the canonical variables of LQG are loops rather than metric fields at points, no physical content is encoded in the values of the metric at specific spacetime points, and therefore we will never have any need to identify the same point across different branches.[17] However, this looks like a difficult option to implement within an EPS-style scheme: the EPS scheme is originally intended as an approach to reducing *local* kinematics to empirical observations on light and (free) particle movement—a kinematics that is not local is thus by definition out of its reach.

---

[15]See nLab authors (2022).

[16]The move by Westman and Sonego (2008) to construct scalars and use them to span a new, diffeomorphism-invariant 'beable' space, and Earman's 1977 notion of Einstein algebra might be suitable (Kretschman-objection resistant) generally covariant alternatives to the metric as well. Notably, in particular for the Einstein/Leibniz algebra approach it is still controversial, though, as to whether this approach does not just suffer from an analogous issue with diffeomorphism invariance.

[17]See for instance Rovelli (1991), in particular §1.5, for a definition of the loop variables.

## 4 Variants of Quantum EPS

Following the discussion in previous sections, we now propose several different ways in which one might approach a quantum EPS construction. None of these options is necessarily right or wrong: there is simply a choice to be made about how much structure we want to put into the kinematics versus the dynamics, and about what exactly we mean by 'spacetime' in the first place. Importantly, all these proposals will have to circumvent the obstacles regarding the identities of spacetime points which were raised in the previous section. In fact, each variant we present in the following is inspired by one of those proposals (1)-(3) for dealing with these obstacles. More precisely, in §4.1 we suggest a strategy which proceeds by ruling out the existence of superpositions of spacetimes (like Penrose's approach); in §4.2 we discuss a strategy based on non-local features (like Barbour's approach); and in §4.3 we suggest an approach which defers most of these matters to the dynamics (like Hardy's approach).

### 4.1 Option (i): Operationalism about spacetime

One of the lessons of the foregoing discussion is that gaining operational access to different branches of a spacetime superposition is extremely nontrivial. As a result, one might start to question whether it makes sense to expect the operationally-motivated axioms of EPS to obtain within each individual branch of the spacetime, given that not all of these branches can have operational significance.

Motivated by this, another possible way of thinking about quantum gravity kinematics would be really to commit to the operational approach and argue that it has the consequence that we can't have 'spacetime superpositions' at all. That is, spacetime is to be understood as an emergent structure defined in terms of what is directly operationally accessible to us,[18] and thus since superpositions of spacetimes can be detected only indirectly (e.g. via measurements of entanglement as in the BMV experiment), they do not in any concrete sense involve 'real' spacetimes.

This emphasizes the fact that there are two different possible ways of thinking about the emergence of spacetime within quantum gravity. Many current approaches to quantum gravity agree that spacetime as we experience it should be understood as arising from an underlying substratum of quantum 'stuff' which is not defined on any spacetime. There are, however, two possible routes that this could take: we could imagine that first a 'quantum spacetime' (which can participate in superpositions) arises out of the substratum, and then our ordinary single-valued spacetime emerges from the quantum spacetime in the macroscopic limit, or we could imagine that our ordinary single-valued spacetime arises directly out of the substratum, with—at least for all effective purposes—no intermediate layer of 'quantum spacetime.' The former approach is presupposed in the standard analysis of low-energy quantum gravity tests like the BMV experiment where we are invited to suppose that two different spacetimes are superposed. The latter approach would have the interesting consequence that although gravity is due to quantum structure, we nonetheless never get superpositions of different spacetimes. This approach has its attractions: in particular, it would sidestep the whole question of identity of spacetime points across branches, since 'spacetime points' would only be defined in the macroscopic limit where branches become non-interacting due to decoherence.[19] Indeed, this approach would lead to the conclusion that the decoherence mechanism is an essential feature of the emergence of spacetime, and thus in regimes without decoherence we shouldn't expect to arrive at manifold structures of the kind constructed by EPS.

---

[18]Or, at least characterised to some extent in terms of what is operationally accessible.

The strict operational approach has precedents in previous work on operational features of general relativity. In particular, we recall that the connection between the operational features probed by EPS and the metric field is typically understood in terms of (her version of) the equivalence principle: in Knox's words (Knox, 2013), the equivalence principle "expresses just that fact about our matter theories that must be true if systems formed from appropriate matter are to reflect the structure of the metric field, that is, if phenomenological geometry is to reflect the geometry of the metric field." (p. 350).[20] Knox thus argues that since the equivalence principle can only be defined in an approximate and contextual way, it follows that phenomenological geometry is coarse-grained, because "operationalized reference frames are objects of finite spatial extent and therefore can't perfectly instantiate metric structure." (p. 352) Thus any spacetime defined in terms of phenomenological, operationally accessible geometry will also be coarse-grained, meaning that we will not be able to think of such a spacetime "as a background manifold with exact geometrical properties." If we accept Knox's argument, then it follows that the operational axioms used by EPS should not be regarded as exactly true even in the classical case: thus in both the classical case and the quantum case the axioms can be understood as characterising high-level emergent features, which means that superpositions of spacetimes need not ever appear.

Of course, if we define spacetime in this operational way we still need some mathematical way of describing what goes on in scenarios like the BMV experiment, but in principle this can always be done using an effective field theory description. That is, we define a 'background spacetime' using directly observable operational procedures: since we never observe superpositions directly, this background spacetime is necessarily classical and single-valued. Then small fluctuations in the metric can be modelled as a quantum field defined on top of this classical, single-valued spacetime: thus the fluctuations can be in superpositions but the background spacetime itself cannot. This is how the BMV experiment is currently modelled—it is taken for granted that we have a background laboratory frame which is used to define meaningful mappings between the branches of the 'spacetime superposition.'

---

[19]Such an approach is realised in various emergent gravity approaches. For a critical survey, see Linnemann and Visser (2018).

[20]For further discussion on the coincidence of (a) what is measured operationally via material fields and (b) geometrical structure, see Read et al. (2018).

That said, the really difficult regimes for quantum gravity are those in which no effective field theory description is possible. Taking a hard-line operational approach to spacetime suggests that, in fact, an effective field theory description is always possible, because the operationally defined spacetime must always exist and be single-valued. Problems might arise, however, when we get into regimes where the behaviour of the background spacetime can't be correctly modelled without taking into account quantum gravity effects.

What would this approach mean for a quantum EPS construction? Straightforwardly, it would imply that the quantum EPS construction would just reduce to the classical EPS construction, because all the same classical axioms will be true of spacetime at the operationally accessible level, and we would not expect the underlying non-operational structure to be captured by an EPS-style axiomatisation.

## 4.2 Option (ii): Non-local scenario

A second possible approach to a quantum version of EPS involves giving up the ambition of a *local* constructive approach to quantum gravity, and instead building up our spacetime from non-local objects like loops, or other objects which are guaranteed to be preserved under (q-)diffeomorphisms. Indeed, this is exactly the approach that has been taken with some success in the loop quantum gravity programme (see for instance Rovelli (2004); notably, even the gravitational path integral can be understood in this way, cf. Rovelli and Vidotto (2015)).

Clearly this option amounts to a severe departure from classical EPS' core principle of constructiveness, insofar as that is understood as involving a commitment to locality. Thus acceptance of this option could be interpreted as an acknowledgement that 'locality' in the usual sense simply can't be preserved within quantum gravity. However, recall that we noted in the introduction that one can simply read EPS as an ontological-relationist program which builds up spacetime from what count naïvely as high-level notions. On this construal, not all 'high-level' notions need be operationally or in any other sense immediately accessible; so one could still accept a 'quantum EPS' based on loops or other such structures as a legitimate extrapolation of the original methodology to the quantum domain.

Indeed, a non-local construction can be achieved quite naturally within the structures postulated by classical EPS. For we may simply include in our axioms a postulate ensuring that superpositions always or at least sometimes come to an end (i.e. the different copies of a light ray in different branches of the superposition ultimately meet and are combined back into a single light ray). We do not have to commit here to an account of how this comes about—that can be added at the dynamical stage, e.g. by best matching. Thus pairs of manifolds should be defined such that they coincide at all times except inside one or more causal diamonds, where a causal diamond is the intersection of the causal future of some spacetime point $p_1$ and the causal past of some other spacetime point $p_2$ (for example, $p_1$ may be identified as the point at which some mass first enters a spatial superposition and $p_2$ may be the point at which the spatial branches of the mass are recombined).

This route fits very nicely into the EPS approach, because an 'echo' as employed by EPS is exactly the right kind of structure to define a causal diamond. An 'echo' in the EPS construction is a map from a point $P$ on the trajectory of a particle to a later point $Q$ on the same trajectory, understood to be defined by sending a light ray away from the first point, and then receiving it back at the later point after its reflection from some object. The path taken by the light ray in this case defines precisely one of the boundaries of the causal diamond defined by $P$ and $Q$. Echoes play a significant role in the EPS construction, particularly in the development of radar coordinates, and thus in effect the structure we need to implement a nonlocal quantum EPS is already inherent in classical EPS. Thus, motivated by this construction, one could imagine formulating an EPS scheme which takes as its basic object not 'messages' but 'causal diamonds,' which ensures that spacetime superpositions are 'anchored' in a time-symmetric way such that the resulting manifolds coincide both in the past and in the future. If we take this approach then we will not be postulating a set of distinct manifolds as in Hardy's picture: instead we will have a single manifold which branches at some points and then recombines, i.e. a non-Hausdorff manifold.

Another interesting possibility on the subject of non-locality would be to explore an approach based on the conjecture that spacetime structure can be derived from entanglement, which has recently been popularised within quantum foundations.[21] These proposals are based on the fact that entanglement in certain sorts of systems obeys an 'area law,' which inspires the conjecture that perhaps we can arrive at a spacetime metric by writing the distance between spacetime regions as a function of the degree of entanglement between them. This approach shares with EPS the motivation of constructing spacetime out of simple first principles, but since entanglement is a paradigmatic example of non-locality, this approach will evidently not satisfy the locality criterion. But, if we take route (ii) and embrace a non-local EPS approach, one might hope to find a fruitful unification of the entanglement-based approach and EPS. One important limitation of the entanglement-based approach is that since it defines distances between regions in terms of the total amount of entanglement between those regions, it is quite coarse-grained and thus may not be well-equipped to assign different distances within different branches of the wavefunction—and this may lead to problem in cases like the BMV experiment, since it is very important to the interpretation of this experiment that the particles are indeed at different distances in different branches. But since the EPR construction has been designed to allow us to construct the metric in full detail, it is possible that supplementing the entanglement approach with some EPS-like axioms would give it the capability to deal with these sorts of cases, so a combination of the two constructions could be of great interest even if not precisely in the spirit of the original EPS approach.

---

[21] See *inter alia* Van Raamsdonk (2010), Jaksland (2021), Cao et al. (2017).

### 4.3 Option (iii): Laissez-Faire scenario

This option involves simply deferring all questions about maps between branches and recombination to the dynamics. It is then very straightforward to arrive at a quantum EPS construction: essentially, we replace classical events with quantum events and then simply require that all the same axioms hold individually within each branch of the wavefunction. This setup is minimal, but at least *prima facie* it would seem to provide a sufficiently well-defined arena relative to which dynamics can be set up.

Although this is the most straightforward option, some doubts remain. First, one might worry that the resulting kinematics will be too general to be useful—we might well be better off with a more restricted kinematics which better reflects the constraints. Second, one might also wonder if it is really natural to expect all the classical axioms to hold individually within each branch of the wavefunction, especially since the axioms are operationally defined and it is not straightforward to operationally access distinct branches of the wavefunction.

## 5 Explicit Laissez-Faire Quantum EPS constructions

In this section, we present two possible starting points for a quantum EPS axiomatisation based on the laissez-faire approach presented in §4.3. In the first, we explicitly quantise rays and particles using a Hilbert space built out of events; in the second, we take inspiration from the consistent- and decoherent-history approaches to quantum mechanics (see Griffiths (2019) for an overview) as well as the (not fully unrelated) branching spacetime literature (see Luc (2020); Luc and Placek (2020) for recent introductions). In each case we present the first few classical EPS axioms and alongside offer the corresponding axioms for a quantum EPS construction.

### 5.1 In terms of superpositions

We begin with a laissez-faire approach based upon quantised rays and particles.

1. *A point set $M = \{p, q...\}$ is a set of events.*

    **Q:** A point set $M = \{p, q...\}$ is a set of events, where each event is identified with an element in a quantum basis $\{|v_p\rangle, |v_q\rangle...\}$, so $M$ defines a Hilbert space $\mathbb{M}$ whose dimension is equal to $|M|$.

2. *Light rays and particles are subsets of M.*

   **Q:** Light rays and particles are sets of superpositions of events, e.g. $P = \{\sum_i c_i|v_i\rangle, \sum_i c_i'|v_i\rangle...\}$ where $\sum_i |c_i|^2 = 1$ where $\sum_i |c_i'|^2 = 1$. With each particle $P$ we can associate a Hilbert space $\mathbb{P}$ which is equal to the subspace of $\mathbb{M}$ defined by all and only those events $|e\rangle$ such that some element of $P$ has support on $|e\rangle$. Likewise, with each light ray we can associate a Hilbert space $\mathbb{L}$ defined similarly.

3. *The map $e_Q : P \to P, p \mapsto e_Q(p)$ is called an echo on P from Q.*

   **Q:** The map $e_Q : \mathbb{P} \to \mathbb{P}, \sum_i c_i|v_i\rangle \mapsto e_Q(\sum_i c_i|v_i\rangle)$ is called an *echo* on $P$ from $Q$. The echo map is a map on just the coefficients $\{c_i\}$, i.e. it transforms only coefficients and not events themselves.

4. *The map $m : P \to Q, p \mapsto m(p)$ is called a message from P to Q.*

   **Q:** The map $m : \mathbb{P} \to \mathbb{Q}, \sum_i c_i|v_i\rangle \mapsto m(\sum_i c_i|v_i\rangle)$ is called a *message* from $P$ to $Q$.

5. *Axiom $A_1$: every particle is a smooth, one-dimensional manifold and any echo on P from Q is smooth and smoothly invertible.*

   **QA1:** Every particle can be written as a superposition of smooth, one-dimensional manifolds, i.e. we can write $P = \sum_{c_i} c_i|V_i\rangle$ where each $V_i$ is a set of basis elements which correspond to points that belong to a smooth, one-dimensional manifold. An echo on P from Q can be split into a superposition of echoes which are smooth and smoothly invertible, i.e. when an echo map is applied to a point $\sum_i c_i|v_i\rangle$, the result is $\sum_i c_i|q_i\rangle$ where we can identify a smooth and smoothly invertible echo from $c_i$ to $q_i$ for all $i$. In the case where the echo has a fixed starting point this means we end up with several different smooth manifolds with the same starting point but different middle and endpoints.

6. *Axiom $A_2$: any message from a particle P to another particle Q is smooth.*

   **QA2:** We require that any message from a particle $P$ to another particle $Q$ can be split into a superposition of messages which are smooth.

7. *Axiom $A_3$: there exists a collection of triplets $(U, P, P')$ where $U \subset M$ and $P, P' \in \mathcal{P}$ [with $\mathcal{P}$ the set of particles] such that the system of maps $\{x_{PP'|U}\}$ is a smooth atlas for M. Each map is written in terms of coordinates $(u, v, u', v')$ where u and v are emission and arrival times at P and likewise on P'.*

   **QA3:** There exists a collection of triplets $(U, P, P')$ where $\mathbb{U} \subset \mathbb{M}$ and $\mathbb{P}, \mathbb{P}'$ such that the system of maps $\{x_{PP'|U}\}$ is a smooth atlas for $M$. Each map is written in terms of coordinates $(u, v, u', v')$ where $u$ and $v$ are emission and arrival times at $P$ and likewise on $P'$.

8. *Claim: Every particle is a smooth curve in M.*

   **QC:** The proof of this claim can be rewritten using the axioms above: it follows straightforwardly from linearity that every particle is a superposition of smooth curves in *M*.

9. *Axiom $A_4$: Every light ray is a smooth curve in M.*

   **QA4:** Every light ray is a superposition of smooth curves in *M*.

10. *Axiom $L_1$: any event e has a neighbourhood V such that each event p in V can be connected within V to a particle by at most two light rays. Given such a neighbourhood and a particle P through e, there is another neighbourhood $U \subset V$ such that any event p in U can be connected with P within V by precisely two light rays and these intersect P in two distinct events $e_1, e_2$. If t is a coordinate function on $P \cap V$ with $t(e) = 0$, then $g = -t(e_1)t(e_2)$ is a function of class $C^2$ on U (i.e. it is twice differentiable on U).*

    **QL1:** Because in our construction each event occurs only in a single branch of the wavefunction, for each event we can define a neighbourhood *V* of points that are within the same branch, and such that each event $p \in V$ can be connected within *V* to a particle by at most two light rays. Given such a neighbourhood and a particle *P* through *e*, there is another neighbourhood $U \subset V$ such that any event $p \in U$ can be connected with *P* within *V* by precisely two light rays and these intersect *P* in two distinct events $e_1, e_2$ (and since *U* is included in *V* it also belongs to the same branch of the wavefunction). If *t* is a coordinate function on $P \cap V$ with $t(e) = 0$, then $g = -t(e_1)t(e_2)$ is a function of class $C^2$ on *U* (i.e. it is twice differentiable on *U*).

11. *Axiom $L_2$: the set $L_e$ of light-directions at an arbitrary event e separates the projective space at e into two connected components. In the tangent space at e, the set of all non-vanishing vectors that are tangent to light rays consists of two connected components.*

    **QL2:** the set $L_e$ of light-directions at an arbitrary event *e* separates the projective space at *e* into two connected components. In the tangent space at *e*, the set of all non-vanishing vectors that are tangent to light rays consists of two connected components. (Note that all of this will take place within the same branch of the wavefunction: no relationships between the projective spaces for different branches can be postulated in the laissez-faire approach).

12. Now we explore the properties of *g* and extract a rank-two tensor from it.
    (a) $\partial_\mu g = 0$.
    (b) $g_{\mu\nu} = \partial_\mu \partial_\nu g$ defines a tensor at *e*.
    (c) The tangent vector $T_\mu$ of any light ray *L* through the point *e* satisfies $g_{\mu\nu} T_\mu T_\nu = 0$.

(d) $g_{\mu\nu} \neq 0$ on particle rays.

**Q:** Since we have defined $g$ using only points which lie in the same branch of the wavefunction as the original point $e$, this part of the construction is exactly the same as the classical case and thus we obtain a rank-two tensor defined on a local 'patch' all within a single branch of the wavefunction.

We will not go through the remaining EPS axioms explicitly here, because now that we have set up the superposition of manifolds and their coordinatisations, the remaining axioms do not have to be altered from the classical case.

## 5.2 In terms of branching spacetime

We turn now to another possible laissez-faire approach, this time based upon the notion of branching spacetimes.[22] The definitions of the set of events, particles, and light rays, as well as of messages and echoes, are similar to their counterparts in the classical EPS approach, the only difference being that the maps corresponding to messages and echoes in the quantum case have to be allowed to be multi-valued.

1. A point set $M = \{p, q...\}$ is a set of events.

2. Light rays and particles are subsets of $M$.

3. The (possibly multi-valued) map $e_Q : P \to P$, $p \to e_Q(p)$ is called an *echo* on $P$ from $Q$.

4. The (possibly multi-valued) map $m : P \to Q$, $p \to m(p)$ is called a *message* from $P$ to $Q$.

We then add a new axiom on branching structure:

5. **QA0:** Each particle (set) $P$ can be decomposed as $P = \bigcup_{i=0}^{n} P_i$ with $P_i = P_{i-1} \cup \bigcup_{j=1}^{m_i} p_{j_i}$ and $P_0, p_{j_i}$ such that messages from other particles onto $P_0, p_{j_i}$ are always single-valued. The subsets of $P$, $B_{j_1,...,j_n} := P_0 \cup \bigcup_{i=1}^{n} p_{j_i}$ are called *particle branches* of $P$.

The standard axioms need to be understood without any assumption that the resulting manifold be Hausdorff (as assumed tacitly in the original EPS construction: see Linnemann and Read (2021a)):

6. *Axiom $A_1$: every particle is a smooth, one-dimensional manifold and any echo on $P$ from $Q$ is smooth and smoothly invertible.*

   **QA1:** Every particle is a one-dimensional non-Hausdorff manifold. Any echo on a branch of $P$, denoted by $P_B$, from a branch of $Q$, denoted by $B_Q$, is smooth and smoothly invertible.

---

[22]For a recent book-length introduction to branching spacetimes and their philosophical significance, see Belnap et al. (2022).

7. *Axiom $A_2$: any message from a particle P to another particle Q is smooth.*

   **QA2:** Any message from a particle branch $B_P$ to another particle branch $B_Q$ is smooth.

8. *Axiom $A_3$: there exists a collection of triplets $(U, P, P')$ where $U \subset M$ and $P, P' \in \mathcal{P}$ [with $\mathcal{P}$ the set of particles] such that the system of maps $\{x_{PP'|U}\}$ is a smooth atlas for M. Each map is written in terms of coordinates $(u, v, u', v')$ where u and v are emission and arrival times at P and likewise on P'.*

   **QA3:** There exists a collection of triplets $(U, B_P, B_{P'})$ where $U \subset M$ and $B_P, B_{P'}$ are particle branches relative to $P$ and $P'$ such that the system of maps $\{x_{B_P B_{P'}|U}\}$ is a smooth atlas for $M$. Each map is written in terms of coordinates $(u, v, u', v')$ where $u$ and $v$ are emission and arrival times at $B_P$ and likewise on $B_{P'}$.

9. *Claim: Every particle is a smooth curve in M.*

   **QC4:** Every particle branch $B_{(j_1,...j_n)}$ is a smooth curve in $M$ with $C : [0,1] \to M$.

   *Proof:* See Linnemann and Read (2021a) for the proof in the case of $M$ being Hausdorff; the proof, however, does not depend on the Hausdorff assumption, and thus carries over.

10. The nature of particles can be characterised further as follows:

    **QA4:** Consider a particle $P$: it is the union of all its branches, i.e. $P = \bigcup_{(j_1,...,j_n)} B_{(j_1,...,j_n)}$; each branch $B_{(j_1,...,j_n)}$ is denoted by the branch index $(j_1,...,j_n)$ and has a corresponding curve $C_{(j_1,...,j_n)} : [0, 1) \to M$. Then for each such curve $C_{(j_1,...,j_n)}$, there exists:

    (i) $g \in (0, 1)$, and
    (ii) a branch index $(j_1,...j_{i-1}, j'_i,...,j'_n)$ such that $C_{(j_1,...,j_n)} = C_{(j_1,...j_{i-1}j'_i,...,j'_n)}$ on $[0, g)$ and $C_{(j_1,...,j_n)} \neq C_{(j_1,...j_{i-1}j'_i,...,j'_n)}$ on $[g, 1]$.

    In other words, each particle $P$ is a multifurcate curve of the second kind in $M$.[23]

    **Remarks:** (1) The axiom is not genuinely empirical as only single branches of the branching structure can be subject to observation: this point has already been discussed in the previous sections of this article. (2) Importantly, and unlike the original EPS for axiom $A_3$, we do not assume implicitly that the smooth manifold established by **QA3** be Hausdorff. (Note

---

[23]The branching spacetime literature standardly features *bifurcate* curves of the second kind:

> A bifurcate curve of the second kind is a pair of curves $C, C'$ in a manifold $W$, with $C, C' : [0, 1] \to W$, such that $C = C'$ on $[0, g)$ and $C \neq C'$ on $[g, 1]$ for some $g \in (0, 1]$. (Luc, 2020, Definition 5)

The multifurcate curve is its natural generalisation.

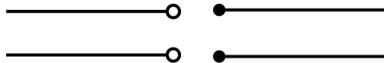

Figure 1: One-dimensional branching structure arising from identifying the two lines on the left hand side.

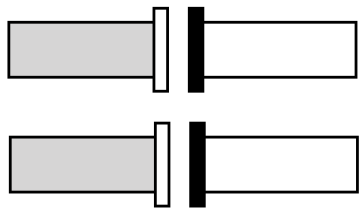

Figure 2: Two-dimensional branching structure arising from identifying the two areas on the left hand side.

that the manifold is therefore not guaranteed to be determined uniquely.) Given that standard differential geometry works with the Hausdorff assumption as a tacit presupposition, it is worth stressing that smooth manifolds can very well be non-Hausdorff. For a first intuitive grasp, consider how figures 1 and 2 illustrate how charts can still be straightforwardly defined on branching lines and surfaces, providing the right intuition as to why this is so on branching (and thus non-Hausdorff) manifolds more generally as well. In particular, one sees from the comparison of the surface to the line case that the idea of a submanifold—effectively a down-projection of a manifold into a lower dimension space, when seen from the right chart—does not depend on the Hausdorff condition. (3) An alternative to (the above formulation of) **QA4** is this: Every particle segment $p_{j_i}$ is a half-open curve $C : [0, 1) \to p_{j_i}$ with $\lim_{a \to 1} C(a) \in p_{j_{i+1}}$ for some $j$.

**QC1:** $P$ is a submanifold of $M$.

*Proof:* This is analogous to part 1 in the proof of the analogous classical claim by Linnemann and Read (2021a).

**QC2:** $P$ is non-Hausdorff submanifold of $M$.

*Proof:* This is an immediate consequence of **QA4**.

**QC3:** $M$ is a non-Hausdorff manifold.

*Proof:* This follows from the contraposition of that submanifolds of Hausdorff manifold are Hausdorff again, and the claims before.

**Remark:** Importantly, a non-Hausdorff differentiable manifold is not uniquely determined by providing atlas structure—after all, it can have both local and global branching. However, the local branching is to be determined through observations of the observer (see below) while the global branching is the kind of fact one can be happy to stay ignorant about in a local approach (akin to how we cannot expect to learn about the global nature of the metric in general relativity).

11. *Axiom $A_4$: Every light ray is a smooth curve in M.*

    **QA4:** Every light ray is a superposition of smooth curves in $M$.

12. *Axiom $L_1$: any event e has a neighbourhood V such that each event p in V can be connected within V to a particle by at most two light rays. Given such a neighbourhood and a particle P through e, there is another neighbourhood $U \subset V$ such that any event p in U can be connected with P within V by precisely two light rays and these intersect P in two distinct events $e_1, e_2$. If t is a coordinate function on $P \cap V$ with $t(e) = 0$, then $g = -t(e_1)t(e_2)$ is a function of class $C^2$ on U (i.e. it is twice differentiable on U).*

    **QL1:** Any event $e$ has a neighbourhood $V$ such that each event $p$ in $V$ can be connected within $V$ to a particle branch by at most two light rays. Given such a neighbourhood and a particle branch $B_P$ through $e$, there is another neighbourhood $U \subset V$ such that any event $p \in U$ can be connected with $B_P$ within $V$ by precisely two light rays and these intersect $B_P$ in two distinct events $e_1, e_2$. If $t$ is a coordinate function on $B_P \cap V$ with $t(e) = 0$, then $g = -t(e_1)t(e_2)$ is a function of class $C^2$ on $U$ (i.e. it is twice differentiable on U).

    **Remark:** One might wonder why messages/echos introduced before are not taken to lead to branching structures. Now, we can assume that light rays *qua* messages/echos are determinate, i.e., not branching themselves, given that the axiom that there is at most one signal going back and forth ($L_1$) for a certain neighbourhood of the particle is arguably still valid even if several signals become possible. For each signal can only differ by direction but—being lightlike—not by speed; and given that the signal has to hit a specific other particle branch from a given particle branch, a restriction from the classical case such as that at most only one echo exists seems to be unchanged by whether or not *several* light signals are actually emitted in different directions (as in the quantum case)—or just hypothetically (as in the classical case).

13. *Axiom $L_2$: the set $L_e$ of light-directions at an arbitrary event e separates the projective space at e into two connected components. In the tangent space at e, the set of all non-vanishing vectors that are tangent to light rays consists of two connected components.*

> **QL2:** the set $L_e$ of light-directions at an arbitrary event *e* separates the projective space at *e* into two connected components. In the tangent space at *e*, the set of all non-vanishing vectors that are tangent to light rays consists of two connected components.
>
> **Q:** Since we have defined *g* using only points which lie in the same branch of the wavefunction as the original point *e*, this part of the construction is exactly the same as the classical case and thus we obtain a rank-two tensor defined on a local 'patch' all within a single branch of the wavefunction.

Just as in the previously-discussed possible laissez-faire construction, we will not go through the remaining EPS axioms explicitly here, because now that we have set up the branching manifold picture together with its coordinatisation, the remaining axioms do not have to be altered from the classical case.

It is worth pointing out that many of the central charges against branching spacetime formulations[24] have little force against the branching spacetime formulation of EPS *specifically*. To see this, consider the two main standard worries that concern (i) a supposed arbitrariness as to when/where the branching occurs, and (ii) violations of energy conservation. The first charge is a serious concern if one hopes to defend branching spacetimes as basic objects of general relativity (*pace* (Luc, 2020)), as the field equations do not determine the splitting behaviour. Indeed, not only is it unclear which branching is to be taken but also—as pointed out before—when and where.[25] However, this is not a problem in our quantum EPS construction since the idea here is simply to provide a kinematical setup relative to which a proper dynamics is still to be formulated, and one may reasonably expect that the dynamics will determine the rules of branching, so there will not be any arbirariness. With respect to the second charge, it is even less clear in the quantum context than in the general relativistic context as to why the satisfaction of energy conditions should be insisted upon.

---

[24]See Earman (2008) for a collection of objections, and Luc (2020); Luc and Placek (2020) for various counters.

[25]To be fair, Luc and Placek (2020) do provide an argument in the later course of their paper as to why branching that is not specified with respect to the when and where can be excluded on physical grounds, namely by dismissing non-Hausdorff manifolds which have bifurcating curves as non-physical. Problematically, though, these are exactly the curves to which the quantum EPS construction adheres. But, again, it is important to distinguish branching spacetime structure as basic objects of general relativity as to of a new theory, i.e., the different motivation for introducing branching spacetime structures: Luc and Placek's idea is that, for a gluing of manifolds to be sound, it should be maximal—otherwise the branching would just be arbitrary (this argument is, arguably, a good way to get around the charges of arbitrariness criticised about Luc (2020) above). Then, indeed, non-Hausdorff manifolds that involve local branching, i.e., bifurcations of the second kind, are excluded just in virtue of a theorem by Hájíček that precisely characterises such manifolds as non-maximal. But, as already explicated above, in the kinematical project of quantum EPS, it is not at all necessary to exclude branching curves to evade full arbitrariness in branching: rather, the arbitrariness in branching can be left to the dynamics, while still allowing for a local branching structure.

## 5.3 Comparison of the approaches

Both of the above laissez-faire approaches come with their own strengths and weaknesses. In the superposition-based approach (SP-EPS from now on), spacetime has already been quantised, with a Hilbert space made up of events, so we have a fully quantum spacetime at the level of the kinematics. By contrast, in the branching-based approach (BS-EPS from now on), spacetime is allowed to branch but has not been quantised, and we therefore arrive at what might be regarded as a classical branching spacetime.

One way to interpret the second approach would be to say it is based on a tacit assumption that spacetime branches are all effectively decohered gravitational histories and thus can be treated as classical objects. From this point of view, BS-EPS seems less general than SP-EPS. On the other hand, BS-EPS could still be quantised at a dynamical level—it naturally suggests an implementation of dynamics in the style of a path-integral formulation, with full flexibility as to how much interference between the different branches is to be considered.[26] Indeed, because BS-EPS does not build quantisation into the kinematics, it offers greater freedom to decide later how to quantise, so in this sense BS-EPS could be regarded as being *more* general than SP-EPS. That said, because SP-EPS has already quantised spacetime, there is less work to do at the level of the dynamics, and one might hope that this would make the dynamics easier to formulate. In SP-EPS, the full branching structure (including the metric structure) formally suggests a promotion to a ket state akin to that of relativistic fields in quantum field theories.

A further interesting question concerns whether or not the two approaches are ultimately equivalent: i.e., is it the case that for any possible choice of dynamics on the SP-EPS kinematics, there is some possible choice of dynamics on the BS-EPS kinematics which will lead to an equivalent theory? The answer to this question may depend on the notion of equivalence one has in mind—it's possible that the two are *empirically* equivalent but structurally different in an important way. If the two are equivalent in the sense that one considers most relevant, then the choice between them comes down to pragmatic considerations about which one offers an easier route to a full quantum gravity theory, but if the two are not equivalent then there could be some fact of the matter about which one is a better fit to reality.[27]

## 6 Lessons from Quantum EPS

To recap: in this article, we have up to this point (i) reviewed the core features of the original EPS construction (§2), (ii) considered in some detail the main conceptual obstacle to a quantum EPS construction—namely, the notion of cross-branch spacetime point identity (§3), (iii) discussed various different ways in which a quantum EPS construction might implemented (§4), and (iv) presented explicitly two such approaches which fall into what we have dubbed the 'laissez-faire' category (§5). In this final section, we turn to considering several conceptual upshots from our investigations into a quantum EPS construction.

---

[26]Another (completely different) way to make use of the kinematical structure provided by BS-EPS is to read the branches as representing Bohmian 'trajectories'. See Tumulka (2005) for a concise demarcation of Bohmian trajectories from the paths in the path-integral formulation.

[27]For an introduction to philosophical issues of theory equivalence, see Weatherall (2019a,b).

## 6.1 Spacetime superpositions versus spatial superpositions

Consider two superficially similar cases: a light ray is put into a superposition of two different paths, and a nanoparticle is put into a superposition of two different paths. The conventional quantum description tells us that in the first case, since the difference between the gravitational fields associated with the different paths taken by the light is negligible (strictly speaking this needn't be correct, because of course electromagnetic fields still have stress-energy content, but let us simply assume that light's contribution to the gravitational field is negligable in what follows) the light goes to two different places in the same spacetime, while in the second case, because the particle is massive enough that the two different paths source significantly different gravitational fields, the massive particles end up in two different spacetimes.

However, our quantum EPS construction offers a different perspective. For suppose we take seriously the notion that points of spacetime have their identities in virtue of a construction akin to radar coordinates. Suppose that a light ray $a$ travels in two opposite directions in two branches of the wavefunction, and that a point $p$ is labelled in one branch of the wavefunction as the points where the light rays $a, b$ meet. But $a$ and $b$ do not meet at all in the other branch of the wavefunction, so if we take identities of spacetime points to be derived from radar coordinates it follows that the point $p$ does not even exist in that other branch. The same goes for all other points, and so we conclude that for this kind of structuralist view about the nature of spacetime point identities, even in the case where there are no differences in spacetime structure between the branches, it nonetheless follows that different branches of the wavefunction represent different spacetimes. From this point of view, the case of spatial superpositions and spacetime superpositions are not really as different as they may first appear.

This is important for the EPS kinematics. *Prima facie*, one might think that we would need to distinguish somehow between the two possibilities: ordinary spatial superpositions would correspond to light rays all staying within a single spacetime, while spacetime superpositions would correspond to light rays going off into different spacetimes (or different branches of a single non-Hausdorff spacetime). But the above analysis suggests that actually we don't need to do this: any time one of our light rays splits into two separate paths those paths are to be regarded as belonging to different spacetimes (or different branches of a single non-Hausdorff spacetime). In some cases, there will be different arrangements of matter in those spacetimes and thus we will get two distinct spacetime structures, giving us a spacetime superposition, while in other cases the arrangements of matter will be the same so we will have identical spacetime structures, but since the coupling between matter and spacetime structure comes in only at the level of the dynamics, we don't have to worry about these differences in our quantum EPS construction: kinematically speaking the construction of spacetime is the same in each case.

Thus, on this kind of view about the nature of spacetime point identities, even when we are dealing entirely with ordinary superpositions with no different spacetime structures involved, the problem raised in §3 still comes into play: empirically it has been shown many times that branches of superpositions like this can indeed be recombined, but how is the map between the two branches effected? A naïve substantivalist of course can simply insist that both branches are defined on the same background spacetime, but proponents of the structuralist-type view under consideration here don't have that option, so how can they make sense of interactions between branches of a superposition at all?

One possible approach would be to deny that spacetime is constructed entirely out of the local behaviour of *actual* particles and light rays, and instead identify spacetime with the metric field as constructed out of *hypothetical* particles and light rays.[28] In this case, the direction in which the actual ray of light were to go would not matter, because the structure of the spacetime would come from the fact that hypothetically it could have gone either way: we'll get exactly the same hypothetical possibilities for the passage of light rays in both branches, so both branches of the superposition can be associated with the same background spacetime, which tells us how to map between the two branches. However, one might question whether this modal take on spacetime and metric structure is strictly in accordance with the operationalist and/or relationalist underpinnings of EPS.

Note also that if we say that the existence of identical (modal) spacetime structure is essential to the possibility of mapping between the branches in the spatial superposition case, then it would seem to follow that there can be no such map in the spacetime superposition case, where we no longer have identical spacetime structure. Thus, if we want to maintain the existence of spacetime superpositions (and the possibility of recombination for such superpositions), we can't insist that the identical structure plays any special role here; and in any case since spatial superpositions can be regarded as simply examples of spacetime superpositions in the limit as some measure of difference of structure goes to zero, one might naturally hope that the procedure for mapping points between branches should be roughly the same in each case—so we should use something like best-matching even in the spatial superposition case (we suggest that Barbour himself would probably advocate this).

---

[28]This would be something akin to 'modal relationalism'—see Belot (2011).

## 6.2 Comparison with other cases for the quantum nature of gravity

In the process of constructing a quantum version of EPS, we saw that by starting with some operationally legitimate assumption about quantum signals, we arrive at the conclusion that spacetime structure involves a superposition of spacetimes in one form or the other. In other words, we have thus arrived at yet another plausibility argument as to why gravity should be quantum. In this section we would like to consider how far the plausibility argument for the quantum nature of gravity differs from by-now familiar ones, which are, *inter alia*, arguments from (i) analogy (say to electrodynamics), (ii) inconsistency of semi-classical gravity, and (iii) inconsistency of classical gravity with gravitational-induced entanglement.

More specifically, the quantum EPS argument (QEA) argument goes as follows. Let us understand the EPS construction as a realisation of the relationalist vision of constructing spacetime out of the behaviour of light and matter; then we argue that since light and matter do different things inside different branches of the wavefunction, different spacetimes necessarily get constructed inside these different branches, so we get superpositions of spacetimes and therefore gravity must be quantised. That is, the quantisation of gravity follows directly from the supposition that it is legitimate to construct spacetime out of the behaviour of light and particles, as the original EPS construction does.

Of course, like all arguments to the effect that gravity must be quantum, this argument is based on some assumptions: primarily the relationalist assumption that spacetime is in some sense a codification of the behaviour of light and matter. Evidently the assumptions of QEA are more philosophical than the assumptions used in other plausibility arguments, which mostly employ specific physical conjectures—for example, the argument of Eppley and Hannah (1977) (an argument of type iii)) requires the the assumption that nongravitational measurements lead to a wavefunction collapse. This comparison makes it clear that QEA depends on preexisting philosophical prejudices in ways that the other arguments do not, and therefore QEA will not seem compelling to someone who is not inclined towards the relationalist position in the first place. On the other hand, because QEA argues for the quantization of spacetime purely on the basis of a conviction about the nature of spacetime, it is less vulnerable to refutation by the proposal of alternative models which don't have the physical features which led to inconsistencies in the original model (as, for example, has happened in the case of the Eppley-Hannah argument(Huggett and Callender, 2001; Mattingly, 2006)). So for those who do favour the relationalist view, QEA may seem a more robust argument than some of the alternatives.

QEA also stands out in virtue of the fact that it is kinematical and constructivist, whereas most of the other arguments for the quantum nature of gravity take place at the level of the dynamics. This, again, makes QEA more robust than some alternatives, as it does not depend on any details of the dynamics and thus will be valid for a large variety of choices for the dynamics. However, in virtue of being kinematical the QEA does run into all of the ambiguity is-sues that we have discussed in this article. But if the kinematicist line of the QEA is given up, it is hard to see how one can still be constructivist, which makes up much of the bite of the QEA. To understand the situation better, it is helpful to recall that the classical EPS scheme can, in virtue of its (more or less consequent) (empirical) constructivist nature, show compellingly at what point the phenomena allow for a much more general kinematical arena than Lorentzian geometry—think for instance of how neither torsion nor differentiability of the metric are backed up well from data or other principles, leaving theories such as teleparallel gravity and Finsler spacetime as live options. Similarly then, the quantum EPS scheme, by being constructivist, despite—or perhaps *because of*—its difficulties, makes us aware of the intricacies of formulating quantum general relativity and neighbouring theories. So, as long as we do not think of the constructivist undertaking in any strict epistemic or metaphysical fashion but more as an auxiliary tool for theory construction, we can accept issues of circularity and happily appreciate instances of ambiguities (such as they arise with the kinematicist commitment).

Finally, it is worth comparing the scope of the arguments: QEA establishes the quantum nature of gravity in relation to relatively high-level quantum particles or 'light rays' that themselves do not have internal structure: the treatment of light just happens at the level of ray optics. In a sense, maybe this limitation is a virtue, as the QEA can thus be seen as an argument for quantum gravity even at rather low energies, such as those studied in low-energy perturbative quantum gravity. Note in this context that arguments of type (i) to (iii) also only establish the quantum nature of gravity for a limited scope; in particular, they say little about what quantum gravity (if one still even would like to call it such) would look like at lower energies.[29]

## 6.3 The quantum hole argument

We close by considering how the issues we have discussed in this article may give new insight into the hole argument. To motivate this, we recall Penrose's argument against spacetime superpositions which was introduced in §3.2.1, and we suppose for the moment that Penrose is correct: that is, *if* there exists no privileged mapping of one superposed spacetime to the other, then no time evolution operator can be defined and therefore the superposition must undergo a gravitationally induced collapse. On the other hand, if spacetime points have primitive identities, then there does exist a privileged mapping of one superposed spacetime to the other, and then in turn a time evolution operator *can* be defined and therefore presumably we will not observe any gravitationally induced collapse. If this argument were correct, then this would be an instance where the existence or nonexistence of primitive identities for spacetime points would have an immediate and in principle observable impact on real physics:

---

[29]For instance, with respect to (i), the analogies to electrodynamics, as well as the identification of gravity by analogy as yet another spin-$n$, *prima facie* only establish a quantum theory of gravity in line with (non-fundamental) quantum electrodynamics and (quantum) effective field theory, respectively.

the existence of primitive identities gives us a stable state, whereas the absence of primitive identities leads to instability and collapse.

Now we don't in fact think Penrose's argument is correct, because we agree with Giacomini and Brukner (2021) that there is no reason to think that a spacetime superposition must contain a unique time evolution operator which applies across all the branches of the superposition. However, Penrose's argument nonetheless illustrates the fact that questions around the identity of spacetime points take on a new relevance when we begin to consider the possibility of quantum diffeomorphisms (as introduced in §3). The usual form of the hole argument requires us to make comparisons across different possible worlds, and thus whilst these sorts of arguments may have some indirect empirical consequences (e.g. they may motivate us to impose a diffeomorphism constraint) there is of course no possibility of directly observing relationships between different possible worlds. On the other hand, we can certainly devise observations which 'observe' happenings in different branches of the wavefunction, or tell us something about the relationships between different branches of the wavefunction, so if spacetime superpositions exist then questions about the identities of points across branches may in principle have direct significance for real observations. So the upshot of our discussion here is that introducing quantum phenomena is likely to make hole argument-style concerns more pressing and empirically relevant: the indeterminism at the centre of the argument becomes not merely a matter of which points instantiate which field values, but rather a matter of there being empirically discernible differences.[30]

But the approach we have adopted here also suggests a way of avoiding the hole problem. For in the EPS context, one might hope that the use of radar coordinates should prevent the hole argument from being made. For radar coordinates label spacetime points in terms of their relation to particles and light rays, and since particles and light rays are ultimately made out of fields, presumably when we perform a diffeomorphism on all the fields as in the hole argument, those particles and light rays will be moved along with the diffeomorphism, meaning that the coordinates will also be moved by the diffeomorphism, and therefore the relation between the fields and the coordinates will not change during the diffeomorphism. Thus, it might seem that the radar coordinate approach is a particularly appropriate way of assigning coordinates in the context of general covariance—not only the physics but even the coordinatisation will be invariant under diffeomorphisms. However, this strategy will work only if the radar coordinates are defined with respect to the behaviour of *actual* particles and light rays. If on the other hand we regard the particles and light rays as purely mathematical object which are used to derive some abstract choice of atlas for the point set, then there is no reason why these radar coordinates should be moved when we apply a diffeomorphism to the actual fields defined on that point set, and therefore diffeomorphisms will once again give rise to the hole problem. So if we want to use the radar coordinate scheme as a novel way of avoiding the hole problem, it is important that the light rays and particles

---

[30]Cf. Pooley and Read (2021).

should be actual.

There are a number of obstacles to this way of avoiding the hole problem. First, in order for actual light rays and particles to give rise to a complete, manifold-like coordinatisation of the point set it seems that both particles and light rays would have to be fairly ubiquitous, in order that every point should be reached by enough light rays to ensure that it gets radar coordinates. It's possible that our actual universe does indeed have a sufficiently large number of particles and light rays do to this (e.g. if we make use of the cosmic microwave background) but on the other hand this doesn't seem guaranteed. And making this requirement imposes a constraint for the electromagnetic matter sector from which light rays derive: as light rays result from the high-frequency limit of propagating electromagnetic waves, electromagnetic field tensors (and, consequently, stress-energy-momentum tensors) will only be acceptable if they lead to sufficient light rays upon that limit. In particular, a strictly global electrostatic (as opposed to electrodynamic) field content in the universe would not be sufficient to allow for the existence of actual radar coordinates. Thus if this route is adopted, EPS must be expanded to a program for establishing the right kinematics for Einstein-Maxwell theory rather than just GR.

Second, we note that there is a clash between taking the light rays in the radar coordinates to be actual, and the proof strategy of classical EPS: for the proof relies heavily on the notion that the radar coordinates establish a chart on a manifold and that, once they have done so, charts other than radar coordinates can be used. One could perhaps argue that other coordinate systems can still be used as mathematical coordinates even if it is the radar coordinates which are taken seriously as a meaningful physical labelling of the points, but it would need to be shown explicitly that the proof strategy of EPS still goes through under this interpretation.

Finally, there is not just one possible radar coordinate system: the radar coordinates depend on the two particle trajectories with respect to which they are defined, and each set of radar coordinates is of limited scope. But given that a compatibility condition needs to hold between radar coordinate systems, one may suspect that the radar coordinate change should be regarded as a passive transformation as between any coordinate system change on a manifold, thus inducing also a corresponding active transformation which will necessarily revive a hole-argument-scenario (as already alluded to above). However, to just apply the notion of passive and of active transformation as familiar from the manifold context (and criticise the physical reading of radar coordinates in terms of actual light ray signals) amounts to a *petitio principii*. The real question here is whether one can—and, perhaps also, whether one even needs to—formally express the radar coordinate charts in a way that prevents equivocation with standard manifold charts. One proposal could consist of piggybacking on the manifold structure: we could simply acknowledge the reciprocal dependence of the charts on the the fields they are describing. However, what about the one-dimensional line of the observer? The radar coordinates are characterised in terms of the emission and the arrival times, which are—on the standard EPS approach—values in 1-dimensional coordinate charts for the two defining particle of the radar coordinates.[31] Can we even think of these times as dependent on the field content?[32]

# 7 Outlook

This investigation into the possibility of a quantum generalisation of the EPS construction has given rise to many interesting questions. First, since the EPS approach constructs only kinematics, one might naturally ask whether the *dynamics* of a quantum theory of gravity can be constructed on top of the already established-kinematical structure. Recall for a moment the situation in classical EPS, in which the dynamics—the Einstein field equations in the case of GR—is obtainable from a best-system analysis, i.e., read out as the best codification of how the kinematical structure evolves in different contexts (say by tracking what kind of energy-momentum content—or even, more directly, what kind of matter field values—correspond to what kind of values of the basic kinematical variables established by EPS). If such a proposal already sounded unrealistic in the classical case, how is it ever to be practically implemented in the quantum case where we can only observe one out of multiple branch structures? After all, in the quantum case, data points have to be collected across different contexts as well as histories within one and the same overall context ('branches'). But note that, at the end of the day, the project of setting up a quantum EPS construction is not one of literally arriving at a proposal for a theory of quantum gravity but one of conceptual clarification—with the major lessons summarised in the previous section.

In any case, it seems necessary to say more on how exactly the dynamics is expected to dispel the aforementioned threat of arbitrariness with respect to when and where exactly curves branch—and when and where not. One simple option is to render gravitational branching as induced by matter branching: we are well familiar with how matter branching comes about (say as to why a beam splits in the interaction with a beam splitter); if the splitting of a manifold is simply—and only—due to matter splitting, and if—as standardly seen—the matter splitting is not arbitrary, then there is no issue with arbitrariness in curve branching anymore either. Admittedly, though, this resolution only works for a lower energy regime of general relativity for which gravity is 'slaved' to that matter theory (see Anastopoulos et al. (2021)). For the actual regime of quantum gravity which is decisively marked by independent gravitational degrees of freedom, the issue will be more cumbersome.

---

[31] Axiom $D_1$ has it that " Every particle is a smooth, one-dimensional manifold".

[32] In fact, how to think of the emission and arrival time is a general issue worth further consideration, even if one takes radar coordinates to be on-par with regular coordinates: what is required, and what exactly is supposed to give rise to the observer's emission and arrival time? In what sense is EPS not after all committed to local clocks?

# A General covariance and quantum diffeomorphisms

As mentioned in the main text, Anandan (1997), Hardy (2019), Giacomini and Brukner (2021), and others have all proposed some variant or other of a principle of 'quantum general covariance', which encodes invariance under quantum diffeomorphisms. Note that these authors seem to be equating diffeomorphism invariance with general covariance, though in fact, following Pooley (2017), there is a case to be made that the two notions should not be regarded as being identical. Some unpacking is thus in order.

Various different definitions have appeared in the literature, but following Pooley (2017), let us say that general covariance is the requirement that "the equations expressing [the] laws are written in a form that holds with respect to all members of a set of coordinate systems that are related by smooth but otherwise arbitrary transformations", while diffeomorphism invariance is the requirement that "if $\langle M, F, D \rangle$ is a solution of the theory, then so is $\langle M, F, d_*D \rangle$ for all diffeomorphisms $d$" (here, $M$ is the background manifold, $F$ are any solution-independent fixed fields on the $M$, $D$ are dynamical fields, and stars indicate push forwards[33]). Famously, as pointed out by Kretschmann, the requirement of general covariance is not really a restriction on possible theories at all: even pre-relativistic theories and special relativity can be put in a generally covariant form by simply turning the background metric into a mathematical object which features explicitly in the laws of the theory.[34]

In contrast, diffeomorphism invariance is a stronger requirement: for example, a generally covariant formulation of special relativity where the Minkowski metric is made explicit in the form of a solution-independent fixed field will fail to be diffeomorphism invariant, since applying a diffeomorphism to the dynamical fields but not the metric will move the fields around relative to the metric and thus will typically give rise to a model which is not dynamically possible. However, in most cases it is still possible to find a way of making a theory invariant under diffeomorphisms: we simply have to turn the fixed fields into dynamical fields. For example, in the case of special relativity we could do this by making the metric a dynamical object governed by an equation requiring that its Riemann curvature tensor is zero everywhere, so now diffeomorphisms will act jointly on the metric and the other fields and will thus take solutions into solutions.

Similar points can be made in the quantum case. By analogy with the classical case, let us say that quantum general covariance is the requirement that

> the equations expressing the laws are written in a form that holds with respect to all members of a set of coordinate systems that are related by arbitrary quantum diffeomorphisms.

---

[33]For more on what we take Pooley to mean by a 'fixed field', see Read (2020).

[34]For more on the history of general covariance and Kretschmann's objection, see Norton (1993). The debate over general covariance has since transformed into a debate over the 'background independence' of general relativity: for further discussion of this issue, see Pooley (2017) and Read (2016).

As in the classical case, this requirement is not very substantive: even if the predictions of a theory depend non-trivially on an identity map between branches, the theory can be still put in a generally covariant form if we turn the identity map between branches into a mathematical object which features explicitly in the laws of the theory. Similarly, let us say that invariance under quantum diffeomorphisms is the requirement that

> if $\langle M, F, D \rangle$ is a solution of the theory, then so too is $\langle M, F, d_*D \rangle$, for all quantum diffeomorphisms $d$.

This is a stronger requirement than quantum general covariance, but again, even models with non-trivial dependence on an identity map between branches could be put in such a form, provided we can find a way of making the identity map at least nominally dynamical by writing it as the solution to some equation.

One way to make the requirement of diffeomorphism invariance more substantive in either the classical or the quantum case is to be more strict about what counts as a dynamical field. For example, inspired by Einstein's action-reaction principle, one might argue that any genuinely dynamical object should not only act but also be acted back upon.[35] Under that stipulation, it would follow that in a special relativistic theory the metric would not really be dynamical, meaning that it must be regarded as a fixed field, and therefore special relativity would fail to exhibit diffeomorphism invariance. Similarly, in the quantum case, if the identity map between branches is independent of the behaviour of the matter in the branches, then it would have to be regarded as a fixed field, so any model where the predictions depend non-trivially on the identity map would fail to exhibit diffeomorphism invariance. Under this construal, in order to have invariance under quantum diffeomorphisms, we would either have to ensure that the identity map is purely a form of gauge, or we would have to make the identity map depend on the configuration of matter—for example, by defining an identity map in terms of similarity of matter content. Another way of thinking about general covariance is due to Barbour (2001), who argues that the true empirical content of general covariance is the manner in which the respective four-dimensional objects are assembled out of three-dimensional constituents. If we accept Barbour's argument, then in fact 'quantum general covariance' should entail not that the theory is invariant under quantum diffeomorphisms, but rather than it uses identities between spacetime points in different branches which are determined by best matching and not by some other absolute structure.[36]

# References

Emily Adlam. Operational theories as structural realism. *arXiv preprint arXiv:2201.09316*, 2022.

---

[35] See Brown and Lehmkuhl (2016) for recent discussion of Einstein's understanding of the action-reaction principle.

[36] For further discussion of this proposal and others, see Read (2016).

Jeeva S Anandan. Classical and quantum physical geometry. In *Potentiality, Entanglement and Passion-at-a-Distance*, pages 31–52. Springer, 1997.

Charis Anastopoulos, Michalis Lagouvardos, and Konstantina Savvidou. Gravitational effects in macroscopic quantum systems: a first-principles analysis. *Classical and Quantum Gravity*, 38(15):155012, 2021.

Jürgen Audretsch and Claus Lämmerzahl. Establishing the riemannian structure of space-time by means of light rays and free matter waves. *Journal of mathematical physics*, 32(8):2099–2105, 1991.

Julian Barbour. On general covariance and best matching. In Craig Callender and Nick Huggett, editors, *Physics Meets Philosophy at the Planck Scale*. 01 2001.

Nuel Belnap, Thomas Müller, and Tomasz Placek. *Branching space-times: theory and applications*. Oxford University Press, 2022.

Gordon Belot. *Geometric possibility*. Oxford University Press, 2011.

Sougato Bose, Anupam Mazumdar, Gavin Morley, Hendrik Ulbricht, Marko Toros, Mauro Pasternostro, Andrew Geraci, Peter Barker, M.S. Kim, and Gerard Milburn. Spin entanglement witness for quantum gravity. *Physical Review Letters*, 119(24), Dec 2017. ISSN 1079-7114. doi: 10.1103/physrevlett.119.240401. URL http://dx.doi.org/10.1103/PhysRevLett.119.240401.

Harvey R. Brown. *Physical Relativity: Space-time structure from a dynamical perspective*. Oxford University Press, 2005.

Harvey R. Brown and Dennis Lehmkuhl. Einstein, the reality of space and the action–reaction principle. In P. Ghose, editor, *Einstein, Tagore and the nature of reality*. Routledge, 2016.

Harvey R. Brown and Oliver Pooley. The origin of the spacetime metric: Bell's Lorentzian pedagogy and its significance in general relativity. In Craig Callender and Nick Huggett, editors, *Physics meets philosophy at the Planck scale*. Cambridge University Press, 2001.

Harvey R. Brown and Oliver Pooley. Minkowski space-time: A glorious non-entity. volume The ontology of spacetime. Elsevier, 2006.

ChunJun Cao, Sean M. Carroll, and Spyridon Michalakis. Space from Hilbert space: Recovering geometry from bulk entanglement. *Phys. Rev. D*, 95:024031, Jan 2017. doi: 10.1103/PhysRevD.95.024031. URL https://link.aps.org/doi/10.1103/PhysRevD.95.024031.

Tian Yu Cao. Prerequisites for a consistent framework of quantum gravity. *Studies in History and Philosophy of Science Part B: Studies in History and Philosophy of Modern Physics*, 32(2):181–204, 2001. ISSN 1355-2198. doi: https://doi.org/10.1016/S1355-2198(01)00003-X. URL https://www.sciencedirect.com/science/article/pii/S135521980100003X. Spacetime, Fields and Understanding: Persepectives on Quantum Field.

Rudolf Carnap. *The Logical Structure of the World*. Berkeley-Los Angeles, Univ, 1967.

Lu Chen and Tobias Fritz. An algebraic approach to physical fields. *Studies in History and Philosophy of Science Part A*, 89:188–201, 2021. ISSN 0039-3681. doi: https://doi.org/10.1016/j.shpsa.2021.08.011. URL https://www.sciencedirect.com/science/article/pii/S0039368121001278.

Marios Christodoulou and Carlo Rovelli. On the possibility of laboratory evidence for quantum superposition of geometries. *Physics Letters B*, 792:64–68, 2019. ISSN 0370-2693. doi: https://doi.org/10.1016/j.physletb.2019.03.015. URL https://www.sciencedirect.com/science/article/pii/S0370269319301698.

Erik Curiel. Kinematics, dynamics, and the structure of physical theory. *arXiv: History and Philosophy of Physics*, 2016.

John Earman. Leibnizian space-times and Leibnizian algebras. In *Historical and philosophical dimensions of logic, methodology and philosophy of science*, pages 93–112. Springer, 1977.

John Earman. Pruning some branches from "branching spacetimes". *Philosophy and Foundations of Physics*, 4:187–205, 2008.

J. Ehlers, F.A.E. Pirani, and A. Schild. Republication of: The geometry of free fall and light propagation. *General Relativy and Gravity*, 44:1587–1609, 2012.

Kenneth Eppley and Eric Hannah. The necessity of quantizing the gravitational field. *Foundations of Physics*, 7(1-2):51–68, 1977. doi: 10.1007/bf00715241.

Michael Esfeld and Dirk-Andre Deckert. *A Minimalist Ontology of the Natural World*. Routledge, 2017.

Michael Esfeld and Vincent Lam. Moderate structural realism about space-time. *Synthese*, 160(1):27–46, 2008. doi: 10.1007/s11229-006-9076-2. URL https://doi.org/10.1007/s11229-006-9076-2.

Flaminia Giacomini and Časlav Brukner. Quantum superposition of spacetimes obeys Einstein's equivalence principle, 2021.

Flaminia Giacomini, Esteban Castro-Ruiz, and Časlav Brukner. Quantum mechanics and the covariance of physical laws in quantum reference frames. *Nature Communications*, 10, 01 2019. doi: 10.1038/s41467-018-08155-0.

Robert B. Griffiths. The Consistent Histories Approach to Quantum Mechanics. In Edward N. Zalta, editor, *The Stanford Encyclopedia of Philosophy*. Metaphysics Research Lab, Stanford University, Summer 2019 edition, 2019.

Lucien Hardy. Implementation of the quantum equivalence principle, 2019.

Nick Huggett. The Regularity Account of Relational Spacetime. *Mind*, 115(457): 41–73, 01 2006. ISSN 0026-4423. doi: 10.1093/mind/fzl041. URL https://doi.org/10.1093/mind/fzl041.

Nick Huggett and Craig Callender. Why quantize gravity (or any other field for that matter)? *Philosophy of Science*, 68(S3):S382–S394, 2001.

Nick Huggett, Niels Linnemann, and Mike Schneider. Quantum gravity in a laboratory? *arXiv preprint arXiv:2205.09013*, 2022.

R. Jaksland. Entanglement as the world-making relation: distance from entanglement. *Synthese*, page 9661–9693, 2021. URL https://doi.org/10.1007/s11229-020-02671-7.

Eleanor Knox. Effective spacetime geometry. *Studies in History and Philosophy of Science Part B: Studies in History and Philosophy of Modern Physics*, 44(3): 346–356, 2013.

Niels Linnemann and James Read. Constructive axiomatics in spacetime physics part i: Walkthrough to the Ehlers-Pirani-Schild axiomatisation. 2021a. URL https://arxiv.org/abs/2112.14063.

Niels Linnemann and James Read. On the status of Newtonian gravitational radiation. *Foundations of Physics*, 51(2):53, 2021b. doi: 10.1007/s10701-021-00453-w. URL https://doi.org/10.1007/s10701-021-00453-w.

Niels Linnemann and Kian Salimkhani. The constructivist's programme and the problem of pregeometry. *arXiv preprint arXiv:2112.09265*, 2021.

Niels Linnemann and Manus R Visser. Hints towards the emergent nature of gravity. *Studies in History and Philosophy of Science Part B: Studies in History and Philosophy of Modern Physics*, 64:1–13, 2018.

Joanna Luc. Generalised manifolds as basic objects of general relativity. *Foundations of Physics*, 50(6):621–643, 2020.

Joanna Luc and Tomasz Placek. Interpreting non-Hausdorff (generalized) manifolds in general relativity. *Philosophy of Science*, 87(1):21–42, 2020.

C. Marletto and V. Vedral. Gravitationally induced entanglement between two massive particles is sufficient evidence of quantum effects in gravity. *Physical Review Letters*, 119(24), Dec 2017. ISSN 1079-7114. doi: 10.1103/physrevlett.119.240402. URL http://dx.doi.org/10.1103/PhysRevLett.119.240402.

James Mattingly. *Is Quantum Gravity Necessary?*, volume 25, pages 322–335. 09 2006. ISBN 978-0-8176-4380-5. doi: 10.1007/0-8176-4454-7_17.

Tushar Menon. Algebraic fields and the dynamical approach to physical geometry. *Philosophy of Science*, 86(5):1273–1283, 2019. doi: 10.1086/705508. URL https://doi.org/10.1086/705508.

Flavio Mercati. *Shape dynamics: relativity and relationalism*. Oxford university press, 2018.

nLab authors. Quantization. http://ncatlab.org/nlab/show/quantization, August 2022. Revision 40.

John D. Norton. General covariance and the foundations of general relativity: eight decades of dispute. *Reports on progress in physics*, 56:791–858, 1993.

John D. Norton. Why constructive relativity fails. *The British journal for the philosophy of science*, 59(4):821–834, 2008.

Roger Penrose. On gravity's role in quantum state reduction. *General Relativity and Gravitation*, 28(5):581–600, 1996. ISSN 1572-9532. doi: 10.1007/BF02105068. URL http://dx.doi.org/10.1007/BF02105068.

Oliver Pooley. Background independence, diffeomorphism invariance, and the meaning of coordinates. In Dennis Lehmkuhl, Gregor Schiemann, and Erhard Scholz, editors, *Towards a theory of spacetime theories*. Birkhäuser, 2017.

Oliver Pooley and James Read. On the mathematics and metaphysics of the hole argument. *The British Journal for the Philosophy of Science*, 0(ja):null, 2021. doi: 10.1086/718274. URL https://doi.org/10.1086/718274.

James Read. *Background independence in classical and quantum gravity*. University of Oxford, 2016.

James Read. Geometrical constructivism and modal relationalism: Further aspects of the dynamical/geometrical debate. *International Studies in the Philosophy of Science*, 33(1):23–41, 2020. doi: 10.1080/02698595.2020.1813530. URL https://doi.org/10.1080/02698595.2020.1813530.

James Read, Harvey R. Brown, and Dennis Lehmkuhl. Two miracles of general relativity. *Studies in History and Philosophy of Science Part B: Studies in History and Philosophy of Modern Physics*, 64:14–25, 2018. ISSN 1355-2198. doi: https://doi.org/10.1016/j.shpsb.2018.03.001. URL https://www.sciencedirect.com/science/article/pii/S1355219817300667.

Hans Reichenbach. *Axiomatisation of the theory of relativity*. University of California Press, 1969.

Carlo Rovelli. Ashtekar formulation of general relativity and loop-space non-perturbative quantum gravity: a report. *Classical and Quantum Gravity*, 8(9):1613, 1991.

Carlo Rovelli. *Quantum Gravity*. Cambridge University Press, 2004.

Carlo Rovelli and Francesca Vidotto. *Covariant loop quantum gravity: an elementary introduction to quantum gravity and spinfoam theory*. Cambridge University Press, 2015.

Robert W. Spekkens. The paradigm of kinematics and dynamics must yield to causal structure. In Anthony Aguirre, Brendan Foster, and Zeeya Merali, editors, *Questioning the Foundations of Physics: Which of our Fundamental Assumptions are Wrong?* Springer, 2015.

Roderich Tumulka. Feynman's path integrals and Bohm's particle paths. *European Journal of Physics*, 26(3):L11, 2005.

Mark Van Raamsdonk. Building up space–time with quantum entanglement. *International Journal of Modern Physics D*, 19(14):2429–2435, 2010. doi: 10.1142/S0218271810018529. URL https://doi.org/10.1142/S0218271810018529.

David Wallace. Who's afraid of coordinate systems? an essay on representation of spacetime structure. *Studies in History and Philosophy of Science Part B: Studies in History and Philosophy of Modern Physics*, 67:125–136, 2019. ISSN 1355-2198. doi: https://doi.org/10.1016/j.shpsb.2017.07.002. URL https://www.sciencedirect.com/science/article/pii/S1355219817300965.

James Owen Weatherall. Part 1: Theoretical equivalence in physics. *Philosophy Compass*, 14(5):e12592, 2019a. doi: https://doi.org/10.1111/phc3.12592. URL https://compass.onlinelibrary.wiley.com/doi/abs/10.1111/phc3.12592. e12592 10.1111/phc3.12592.

James Owen Weatherall. Part 2: Theoretical equivalence in physics. *Philosophy Compass*, 14(5):e12591, 2019b. doi: https://doi.org/10.1111/phc3.12591. URL https://compass.onlinelibrary.wiley.com/doi/abs/10.1111/phc3.12591. e12591 10.1111/phc3.12591.

Hans Westman and Sebastiano Sonego. Events and observables in generally invariant spacetime theories. *Foundations of Physics*, 38(10):908–915, 2008.